%% file: main.tex
\documentclass[alpha-refs]{amsart}

\usepackage{fullpage}
\usepackage{siunitx}
\usepackage{graphicx,psfrag,epsf}
\usepackage{enumerate}
\usepackage{url} 
\usepackage[T1]{fontenc}
\usepackage[utf8]{inputenc}
\usepackage{amsmath,amsfonts,amssymb,amscd,xspace}
\usepackage{caption}
\usepackage[labelformat = empty,position=top]{subcaption}
\usepackage{xypic}
\usepackage[color,matrix,arrow,curve]{xy}
\usepackage{rotating}
\usepackage{tikz}
\usetikzlibrary{decorations.pathreplacing}
\usepackage{adjustbox}
\usepackage{float}
\usetikzlibrary{arrows}
\RequirePackage{algorithmic}
\RequirePackage[section]{algorithm}
\usepackage{booktabs}
\usepackage{pifont}
\usepackage[numbers]{natbib}
\newcommand{\norm}[1]{\left\lVert#1\right\rVert}

\input{definitions.tex}

\begin{document}
\title[Causal Structure Learning with Hidden Variables]{Robust causal structure learning with some hidden variables}

\author{Benjamin Frot}
\thanks{BF and PN are equally contributing authors.}
\address{Seminar f\"ur Statistik, ETH Z\"urich, Z\"urich, Switzerland}

\author{Preetam Nandy}
\address{Department of Biostatistics and Epidemiology,
University of Pennsylvania, Philadelphia, USA}
\author{Marloes H. Maathuis}
\address{Seminar f\"ur Statistik, ETH Z\"urich, Z\"urich, Switzerland}
\curraddr{Marloes H. Maathuis, Seminar f\"ur Statistik, ETH Z\"urich, Z\"urich, Switzerland}
\email{maathuis@stat.math.ethz.ch}
\date{\today}






\begin{abstract}
We introduce a new method to estimate the Markov equivalence class of a directed acyclic graph (DAG) in the presence of hidden variables, in settings
where the underlying DAG among the observed variables is sparse, and there are a few hidden variables that have a direct effect on many of the observed ones.
Building on the so-called low rank plus sparse framework, we suggest a two-stage approach which first removes the effect of the hidden variables, and then estimates the Markov equivalence class of the underlying DAG under the assumption that there are no remaining hidden variables.
This approach is consistent in certain high-dimensional regimes and performs favourably when compared to the state of the art, both in terms of graphical structure recovery and total causal effect estimation.

\end{abstract}

\maketitle

\section{Introduction}\label{section:introduction}
\input{Sections/Introduction.tex}
\section{Preliminaries}
\label{section:preliminaries}
\input{Sections/GraphicalModels.tex}
\input{Sections/Transelliptical.tex}
\section{Problem Statement and Suggested Work}
\label{section:suggest_work}
\input{Sections/Setup.tex}
\input{Sections/SuggestedEstimator.tex}
\input{Sections/PreviousWork.tex}

\section{Theoretical Results}
\label{section:theoretical_results}
\input{Sections/Consistency.tex}
\section{Performances on Simulated Data}
\label{section:simulations}
\input{Sections/Simulations1.tex}

\input{Sections/Simulations2.tex}
\input{Sections/Simulations3.tex}
\section{Applications}
\label{section:application}
\input{Sections/Application1.tex}
\input{Sections/Application2.tex}

\section{Discussion}
\input{Sections/Discussion.tex}

\section*{Acknowledgements}
We are grateful to Wu Jun for numerous insightful comments on our proofs and software. 
We are also indebted to the anonymous reviewers for their suggestions on how to expand to the scope of our paper and for pointing out some inconsistencies. 
Their advice brought about significant changes from which the present
work benefited greatly.


\input{output.bbl}
\end{document}

%% file: definitions.tex
\newtheorem{theorem}{Theorem}[section]

\newtheorem*{assumption*}{Assumption}

\DeclareMathOperator{\skeleton}{skeleton}

\DeclareMathOperator*{\argmin}{arg\,min}

\DeclareMathOperator*{\trace}{Trace}

\newcommand{\circcirc}[2]{\xymatrix@C=1.5em{X_{#1}\ar@{*{\circ}-*{\circ}}[r]&X_{#2}}}
\newcommand{\tailarrow}[2]{\xymatrix@C=1.5em{X_{#1}\ar@{->}[r]&X_{#2}}}
\newcommand{\rtailarrow}[2]{\xymatrix@C=1.5em{X_{#1}&\ar@{->}[l]X_{#2}}}

\newcommand*{\indep}{\!\perp\!\!\!\perp}

\newcommand*{\dsep}{\perp}

\newcommand{\xyO}{\cir<2.3pt>{}}

\newcommand{\tK}{K_{nO}}

\newcommand{\tL}{L_{n}}

\newcommand{\tC}{\mathcal{C}_{nO}}

\newcommand{\tR}{\rho_{nij|U}}

\newcommand{\eK}{\hat{K}_{nO}}

\newcommand{\eL}{\hat{L}_{n}}
\newcommand{\eC}{\hat{\mathcal{C}}_{nO}}

\newcommand{\eR}{\hat{\rho}_{nij|U}}

%% file: Sections/Introduction.tex
The task of learning causal directed acyclic graphs (causal DAGs) arises in many areas of science and engineering. 
In such graphs, nodes represent random variables and edges encode direct causal effects.
The problem of recovering their structure from observational data is challenging and cannot be tackled without making untestable assumptions \citep{Pearl09}.
Among other assumptions, \emph{causal sufficiency} is particularly constraining. 
Briefly, causal sufficiency requires that there be no hidden (or latent) variables that are common causes of two or more observed variables (such hidden variables are often called \emph{confounders}). 
Although causal sufficiency is unrealistic in most applications, many structure causal learning algorithms operate under this assumption (\emph{e.g.} \cite{SpirtesEtAl00,Chickering02,TsamardinosEtAl06,NandyHauserMaathuis16}).
On the other hand, methods allowing for arbitrary hidden structures tend to be overly conservative, recovering only a small subset of the causal effects \citep{SpirtesMeekRichardson95,ColomboEtAl12,ClaassenMooijHeskes13}. 
In the present work, we suggest taking a middle-ground stance on causal sufficiency by allowing hidden variables while imposing some
restrictions on their number and behaviour.
More precisely, we consider settings
where the underlying DAG among the observed variables is sparse, and there are a few hidden variables that have a direct effect on many of the observed ones \citep{ChandrasekaranParriloWillsky12}.
This is an interesting problem for at least two reasons. 

First, these assumptions cover important real-world applications. 
In the context of gene expression data, for example, such confounding occurs due to technical factors or unobserved environmental variables (\emph{e.g.} \cite{Leek2007, Stegle2012, Speed2013}). 
For another example, consider the task of modelling the inverse covariance structure of stock returns \citep[\S 9.5]{Hastie:2015}. \cite{ChandrasekaranParriloWillsky12} showed that a large fraction of the conditional dependencies among stock returns can be explained by a few hidden variables, \emph{e.g.} energy prices.
By applying similar ideas to the modelling of the 
California reservoir network, \cite{taeb2017} were able to infer and 
quantify the effect of external phenomena that have a system-wide 
effect on the network. 

Second, this setting is complementary to the realm of application of popular 
algorithms that do not assume causal sufficiency, such as versions of the Fast Causal Inference algorithm \citep{SpirtesMeekRichardson95, ColomboEtAl12, ClaassenMooijHeskes13}.
Under our assumption that there are a few hidden variables that affect many of the observed ones, most observed variables are conditionally dependent given any subset of the observed variables.
Hence, the underlying so-called \emph{maximal ancestral graph} is expected to be dense which, in turn, implies that very few edges can be oriented (see Figure \ref{fig:example} for an example).
Moreover, learning such dense graphs is computationally demanding.

In this paper, we suggest a two-stage procedure. 
First,  the so-called ``low-rank plus sparse'' approach of \cite{ChandrasekaranParriloWillsky12} is
 applied to the covariance matrix to obtain a pair of positive semi-definite matrices, $(\hat{K}_O, \hat{L})$ say,
describing the estimated inverse covariance matrix between observed variables conditional on the hidden ones ($\hat{K}_O$),
 and the estimated effect of the hidden variables ($\hat{L}$). 
In the second stage, a causal structure learning algorithm which assumes causal sufficiency is 
applied to ${\hat{K}_O}^{-1}$. 
In addition, (joint) total causal effects can be straightforwardly estimated using the (joint-)IDA
algorithm \citep{MaathuisKalischBuehlmann09, Maathuis2010, NandyMaathuisRichardson16}.

The suggested approach is conceptually simple and enjoys many desirable theoretical and computational 
properties. 
We study two versions of our estimator. 
One is based on the sample covariance matrix, as described above, and the other on the sample Kendall correlation 
matrix.
Building on recent work by \cite{Wegkamp2016} and \cite{han2017}, we first establish
the convergence rates of the low-rank plus sparse approach for two families
of distributions -- sub-Gaussian random variables and transelliptical distributions -- 
thus extending previous results which assumed Gaussian distributions.
We then derive conditions and scaling regimes under which our
two stage estimators are consistent. 
Through extensive simulations, we show that our approach outperforms other relevant methods, 
both in terms of graph structure recovery and total causal effect estimation. 
Our main focus being on applicability, we also suggest strategies to select the tuning parameters in various settings and illustrate their performances in simulations. 
Finally, we analyse two datasets.
In our first application, we model the expression levels of the genes responsible for isoprenoid synthesis in \emph{Arabidopsis thaliana} and show
that some of the hidden variables we estimate have a clear biological interpretation.
We also model the expression levels of hundreds of genes expressed in ovarian cancer and assess our results using two external sources of validation. 
Compared to state-of-the-art algorithms, we find our approach to be better at recovering known causal relationships\footnote{The code for our simulations and applications is made available with this paper.}.

%% file: Sections/GraphicalModels.tex
\subsection{Graphical Models Terminology}
\label{graphical_models}

We consider graphs $\mathcal{G} = (X,E)$, where the vertices (or nodes) $X = \{X_1,X_2, \ldots\}$ represent random variables 
and the edges represent relationships between pairs of variables. 
The edges can be either \emph{directed} ($\tailarrow{i}{k}$) or \emph{undirected} ($\circcirc{i}{k}$). 
A \emph{directed graph} can only contain directed edges. An \emph{undirected graph} can only contain undirected edges. A \emph{partially directed graph} may contain both directed and undirected edges. 
The \emph{skeleton} of a partially directed graph $\mathcal{G}$, 
denoted as $\skeleton(\mathcal{G})$, is the undirected graph that results from replacing all directed edges of $\mathcal{G}$ by undirected edges.

Two nodes $X_i$ and $X_k$ are \emph{adjacent} if there is an edge between them. 
If $\tailarrow{i}{k}$, then $X_i$ is a \emph{parent} of $X_k$.
A \emph{path} between $X_i$ and $X_k$ in a graph $\mathcal{G}$ is a sequence of distinct nodes $(X_i,\dots,X_k)$
 such that all pairs of successive nodes in the sequence are adjacent in $\mathcal{G}$. 
 A \emph{directed path} from $X_i$ to $X_k$ is a path between $X_i$ and $X_k$, where all edges are directed towards $X_k$.
 A directed path from $X_i$ to $X_k$ together with $\tailarrow{k}{i}$ forms a \emph{directed cycle}. 
 A graph without directed cycles is  \emph{acyclic}.  
 A graph that is both (partially) directed and acyclic, is a \emph{(partially) directed acyclic graph} or (P)DAG.

A DAG encodes conditional independence relationships via the notion of \textit{d-separation} \cite[see][Def. 1.2.3]{PearlBook09}. 
Several DAGs can encode the same set of d-separations and such DAGs form a \emph{Markov equivalence class}. 
A Markov equivalence class of DAGs can be uniquely represented by a \emph{completed partially directed acyclic graph} (CPDAG), 
which is a PDAG that satisfies the following: $\tailarrow{i}{k}$ in the CPDAG if $\tailarrow{i}{k}$ in every DAG in the Markov equivalence class, and $\circcirc{i}{k}$ in the CPDAG if the Markov equivalence class contains a DAG in which $\tailarrow{i}{k}$ as well as a DAG in which $\rtailarrow{i}{k}$ \citep{VermaPearl90, AnderssonEtAl97}. 
In this sense, the circle marks represent uncertainty about the edge marks.

For $S \subseteq X \setminus \{X_i, X_j\}$, we write $X_i \indep X_j | S$ to denote that $X_i$ and $X_j$ are independent given $S$,
while $X_i \dsep_\mathcal{G} X_j | S$
means that $X_i$ and $X_j$ are d-separated by $S$ in $\mathcal{G}$. 
A DAG $\mathcal{G}$ is a \emph{perfect map} of the joint distribution
 of $X$ if for all $X_i, X_j$ such that $X_i \neq X_j$ and for all $S \subseteq X \setminus \{X_i, X_j \}$, we have $X_i \indep X_j | S \Leftrightarrow X_i \perp_\mathcal{G} X_j | S$. 
 
 
When some variables are unobserved, as is assumed in this paper, complications arise because the class of DAGs is not closed under 
marginalisation. 
Among other factors, this limitation prompted the development of another class of graphical independence models called \emph{maximal ancestral graphs} (MAGs) \citep{RichardsonSpirtes02}. 
 A criterion akin to d-separation makes it possible to read-off independencies of such graphs and, since multiple MAGs can encode the same
 set of conditional independence statements, one usually attempts to recover a \emph{partial ancestral graph} (PAG) which describes a 
 Markov equivalence class of MAGs \citep{AliRichardsonSpirtes09}. 
 Like in CPDAGs, circle marks represent uncertainty about edge marks. In particular, a circle mark occurs in the PAG if the Markov equivalence class contains a MAG in which the edge mark is a tail, and a MAG in which the edge mark is an arrowhead \citep{Zhang08-orientation-rules}.

%% file: Sections/Transelliptical.tex
\subsection{Background on sub-Gaussian Random Variables and Transelliptical
Distributions}

In what follows, we will consider structural equation models with
errors that are either
sub-Gaussian or elliptical.

A random variable is
sub-Gaussian if the tails of its distribution decay at least as fast as the tails of a Gaussian distribution.
Formally, a random variable $X$ is said to be \emph{sub-Gaussian} with parameter $\sigma^2$
if $\mathbb{E}X = 0$ and it satisfies
\[
  \mathbb{E} \exp(tX) \leq \exp(\frac{t^2 \sigma^2}{2}),~\forall t \in \mathbb{R}.
\]
A random vector $X \in \mathbb{R}^p$ is sub-Gaussian with parameter
$\sigma^2$ if $\mathbb{E}X = 0$ and $u^TX$ is sub-Gaussian with parameter $\sigma^2$ for all unit
vectors $u \in \mathbb{R}^p$. 
Important examples of sub-Gaussian random variables are Gaussian random variables,
Bernoulli random variables and, more generally, any bounded random variable. We refer the reader to \cite{Vershynin12} for more results
and definitions about sub-Gaussian random variables, including the notion sub-Gaussian norm.

An elliptical distribution is another extension of the multivariate Gaussian distribution.
For any two random vectors $X$ and $Y$, let $X \stackrel{d}{=} Y$ denote
the fact that $X$ and $Y$ have the same distribution.
Then, a random vector $X \in \mathbb{R}^p$ is said to have an \emph{elliptical distribution} if and only if $X$ has stochastic 
representation $X \stackrel{d}{=} \mu + \xi A U$ (Def. 2.1 in \cite{han2017}).
Here, $\mu \in \mathbb{R}^p$, $k := rank(A)$, $A \in \mathbb{R}^{p\times k}$, $\xi \geq 0$ is a random variable independent of $U$, $U$ is uniformly distributed on the unit sphere in $\mathbb{R}^k$.
Letting $\Sigma := A A^T$, we write $X \sim EC_p(\mu, \Sigma, \xi)$. 
We limit ourselves to those
distributions for which $\mathbb{E}\xi^2 < \infty$, thus guaranteeing
the existence of the covariance matrix which is then equal to 
$\frac{\mathbb{E}\xi^2 }{k} \Sigma$.
Any linear combination of elliptically distributed variates is
still elliptical. More precisely, for $X \sim EC_p(\mu, \Sigma, \xi)$,
$B \in \mathbb{R}^{p' \times p}$ and $v \in \mathbb{R}^{p'}$, we have
$v + BX \sim EC_{p'}(B\mu + v, B \Sigma B^T, \xi)$ (Th. 2.16 of \cite{fang1990}).
Interesting examples of elliptical distributions include the family
of multivariate $t$-distributions (with 3 or more degrees of freedom)
and rank-deficient Gaussians. 
We will however assume that $\Sigma$ is non-singular.

Transelliptical distributions -- or semiparametric elliptical copulas -- extend elliptical distributions in that they allow
for some marginal transformations of the random variables.
A random vector $X = (X_1, \ldots, X_p)^T$ follows a 
\emph{transelliptical distribution} (Def. 2.2 in \cite{han2017}) 
if there exist $p$ strictly
increasing univariate functions $f_1, \ldots, f_p$ such that
\[
 (f_1(X_1), \ldots, f_p(X_p))^T \sim EC_p(0, \Sigma, \xi),~\text{where} 
 ~diag(\Sigma) = I_p~\text{and}~\mathbb{P}(\xi =0)=0.
\]
We write $X \sim TE_p(\Sigma, \xi, f_1, \ldots, f_p)$.
Following the terminology of \cite{Liu2012}, $\Sigma$ is called the
\emph{latent generalised correlation matrix}.
Moreover, the family of transelliptical distributions -- and \emph{a fortiori}
the family of elliptical distributions -- is closed under marginalisation and
conditioning (Lemma 3.1 \cite{Liu2012}), a property which allows the definition
of so-called \emph{transelliptical graphical models}.

%% file: Sections/Setup.tex
\subsection{Setup and Notations}
\label{setup}
Throughout, we assume that we are given $n$ independent, identically distributed (i.i.d.) realisations 
of a partially observed, zero-mean random vector $X = (X_O^T, X_H^T)^T \in \mathbb{R}^{p+h}$, where the variables in $X_O$ are observed while the 
variables in $X_H$ remain hidden.
We consider two distinct settings:
\begin{description}
    \item[(Setting 1)] either $X$ is jointly sub-Gaussian with inverse covariance matrix $K \in \mathbb{R}^{(p+h)\times (p+h)}$,
    and there exists a DAG, $\mathcal{G}_O$ say, which is a perfect map of the distribution of
     $X_O$ conditional on $X_H$. 
    It is assumed that the causal mechanism generating $X_O$
     \emph{conditional} on $X_H$
     is of the form
    \begin{equation}\label{eq:subG_SEM}
        X_O \leftarrow B_O X_O + D^{1/2} \epsilon + \Gamma X_H,~\text{with}~cov(\epsilon) = I_p,~
        D \in \mathbb{R}^{p\times p},\Gamma \in \mathbb{R}^{p\times h}.
    \end{equation}
     We assume that an intervention on observed variables has no effect on the distribution of $X_H$. The non-zero pattern of $B_O$ is determined by the causal DAG $\mathcal{G}_O$.
     Furthermore, $D$ is diagonal and $\epsilon$ is a sub-Gaussian random vector which is 
     independent of $X_H$.
    \item[(Setting 2)] or $X$ is transelliptically distributed 
    according to $TE_{p+h}(K^{-1}, \xi, f)$
    with $f := (f_O^T, f_H^T)^T := (f_1, \ldots, f_p, f_{p+1}, \ldots, f_{p+h})^T$, and 
    there exists a DAG, $\mathcal{G}_O$ say, which is a perfect map of the 
    distribution of $f_O(X_O)$ conditional on $f_H(X_H)$.
    It is assumed that the causal mechanism generating $f_O(X_O)$ conditional on $f_H(X_H)$ is of the form
    \begin{equation}\label{eq:trans_SEM}
        f_O(X_O) \leftarrow B_O f_O(X_O) + D^{1/2} \epsilon + \Gamma f_H(X_H),~\text{with}~cov(\epsilon) = I_p,~
        D \in \mathbb{R}^{p\times p},\Gamma \in \mathbb{R}^{p\times h},
    \end{equation}
We assume that an intervention on observed variables has no effect on the distribution of $X_H$. Here, $B_O$ and $\epsilon$ satisfy the same assumptions as in Setting 1, except that we relax the sub-Gaussian assumption while imposing the assumption that $\epsilon$ is an elliptically distributed random vector.
\end{description}

One possible interpretation of \eqref{eq:subG_SEM} and \eqref{eq:trans_SEM} is that they describe linear structural equation models (SEMs) with correlated errors.
We now look at these settings more closely.
Let $K$ be partitioned as follows
 \[
    K =
    \begin{pmatrix}
        K_O& K_{OH}\\
        K_{HO}& K_H
    \end{pmatrix},
\] 
with $K_O \in \mathbb{R}^{p \times p}$, $K_{OH} \in \mathbb{R}^{p \times h}$, $K_H \in \mathbb{R}^{h \times h}$.
The conditional distribution of $X_O$ given $X_H$ is sub-Gaussian with covariance matrix $K_O^{-1}$ or transelliptical with latent generalised correlation matrix $K_O^{-1}$. We assume that there exists a DAG $\mathcal{G}_O$ which is a perfect map of a sub-Gaussian distribution with covariance matrix $K_O^{-1}$ (Setting 1) or a perfect map of an elliptical distribution with correlation matrix $K_O^{-1}$ (Setting 1). Our goal is to estimate $K_O^{-1}$ and the CPDAG $\mathcal{C}_O$ that represents Markov equivalence class of $\mathcal{G}_O$. These estimates can be used in estimating causal effects between observed variables. In fact, under the causal model described by \eqref{eq:subG_SEM}, one can show that the causal effect of $X_i \in X_O$ on $X_j \in X_O$ equals the regression coefficient of $X_i$ in the linear regression of $X_j$ on $X_i$ and $X_i$'s parents in $\mathcal{G}_O$, computed from $K_O$ (see, for example, Proposition 3.1 of the supplementary material of \cite{NandyMaathuisRichardson16}). Similar result holds for the causal model described by \eqref{eq:trans_SEM}. Hence, estimates of $\mathcal{C}_O$ and $K_O^{-1}$ enable
the estimation of multisets of possible causal effects, via
the (joint-)IDA algorithm \citep{MaathuisKalischBuehlmann09, MaathuisColomboKalischBuehlmann10, NandyMaathuisRichardson16}.

Since $X_H$ is unobserved, we need to estimate $\mathcal{C}_O$ and $K_O^{-1}$ from $n$ i.i.d.\ samples from the marginal distribution of $X_O$. A simple calculation yields for Setting 1 that $X_O$ is sub-Gaussian with
$ cov(X_O) = \left(K_O - K_{OH} K_H^{-1} K_{HO} \right)^{-1}$, and for Setting 2 that  
$X_O \sim TE_p \left(\left(K_O - K_{OH} K_H^{-1} K_{HO} \right)^{-1}, \xi, f_O \right)$ (see, for example, Corollary of Th. 2.16 in \cite{fang1990}).
Setting $L := K_{OH} K_H^{-1} K_{HO}$, we have that $L$ summarises the effect of the hidden variables on the
observed ones. In practice, only $n$ samples from these marginal distributions are observed and 
we let $\hat{\Sigma}_n$ be some generic estimator of $\Sigma := (K_O - L)^{-1}$. For example, $\hat{\Sigma}_n$ could be the sample covariance matrix (Setting 1) or a modified sample Kendall correlation matrix (Setting 2). In what follows, conditions on $K_O$ and $L$ will be given for estimating $K_O$ consistently under Settings 1 and 2. We will then use the estimate of $K_O$ to obtain a consistent estimate $\mathcal{C}_O$ under further assumptions. 

To make Settings 1 and 2 easier to comprehend, consider a set of hidden variables $Z_H$ such that $(X_O^T, Z_H^T)^T$ is generated from an acyclic linear SEM with uncorrelated errors:
\begin{align}\label{eq: linear SEM}
\left( \begin{array}{c} X_O \\ Z_H \end{array} \right)  \leftarrow\left( \begin{array}{cc} W_O & W_{OH} \\ W_{HO} & W_H \end{array} \right)
\left( \begin{array}{c} X_O \\ Z_H \end{array} \right) + \left( \begin{array}{cc} D^{1/2}\epsilon \\ D^{1/2}_{H}\eta \end{array} \right),
\end{align}
where $(\epsilon^T, \eta^T)^T$ is a sub-Gaussian random vector with $cov\left((\epsilon^T, \eta^T)^T\right) = I_{p+h}$. Then it follows from straightforward calculation that $X_O$ satisfies Setting 1 with
$X_H = (I - W_H)^{-1}D^{1/2}_{H}\eta$, $B_O = W_O + W_{OH}(I - W_H)^{-1}W_{HO}$, $\Gamma = W_{OH}$ and $\mathcal{G}_O$ equals the DAG that corresponds to the non-zero entries in $B_O$. A similar result holds for Setting 2 with $f_H(X_H) = (I - W_H)^{-1}D^{1/2}_{H}\eta$. 

If we additionally assume $W_{HO} = 0$ in \eqref{eq: linear SEM}, then we have $B_O = W_O$ and $Z_H$ equals $X_H $ or $f_H(X_H)$. The assumption $W_{HO} = 0$ restricts ourselves to linear SEMs where hidden variables do not have observed parents. From a mathematical point of view, this assumption is not necessary. When it holds, however, a qualitative interpretation of our conditions on $K_O$ and $L$ required for consistently estimating $K_O$ is possible. Namely, that there be few hidden variables with widespread effects and that there be few direct causes of each observed variable.

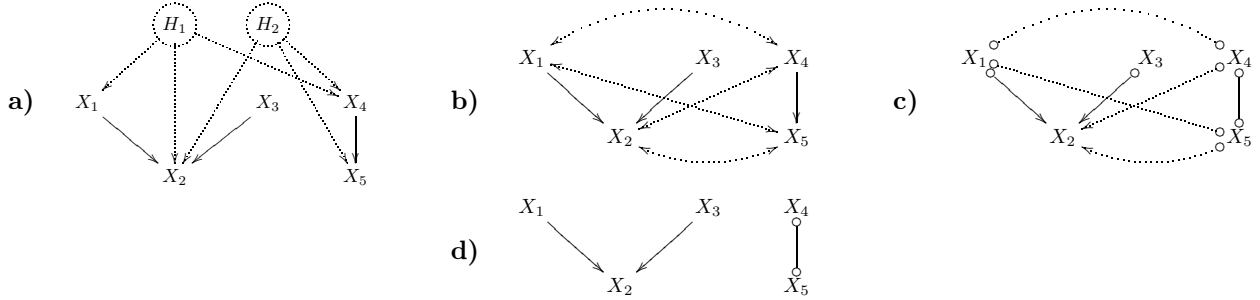
\begin{figure}
    \centering
    \begin{center}
        \begin{subfigure}[t]{0.03\textwidth}
            \textbf{a)}
        \end{subfigure}
        \begin{subfigure}[t]{0.25\textwidth}
            \centering
            \adjustbox{valign=m}{
            \resizebox{\textwidth}{!}{
                \input{Figures/Example/example_DAG.tex}}}
        \end{subfigure}\hfill
        \begin{subfigure}[t]{0.03\textwidth}
            \textbf{b)}
        \end{subfigure}
        \begin{subfigure}[t]{0.25\textwidth}
            \centering
            \adjustbox{valign=m}{
            \resizebox{\textwidth}{!}{
                \input{Figures/Example/example_MAG.tex}}}
        \end{subfigure}\hfill
        \begin{subfigure}[t]{0.03\textwidth}
            \textbf{c)}
        \end{subfigure}
        \begin{subfigure}[t]{0.25\textwidth}
            \centering
            \adjustbox{valign=m}{
            \resizebox{\textwidth}{!}{
                \input{Figures/Example/example_PAG.tex}}}
        \end{subfigure}\\
        \begin{subfigure}[t]{0.03\textwidth}
            \textbf{d)}
        \end{subfigure}
        \begin{subfigure}[t]{0.25\textwidth}
            \centering
            \adjustbox{valign=m}{
            \resizebox{\textwidth}{!}{
                \input{Figures/Example/example_CPDAG.tex}}}
        \end{subfigure}
        \caption{
            An example of a DAG, $\mathcal{G}$, with two hidden variables ($H_1$, $H_2$) and the corresponding constructions.
            \textbf{a)} $\mathcal{G}$.
            \textbf{b)} The MAG associated with $\mathcal{G}$ when $H_1, H_2$ are marginalised out.
            \textbf{c)} The PAG representing the Markov equivalence class of the MAG.
            \textbf{d)} The CPDAG $\mathcal{C}_O$ associated with the observed part of $\mathcal{G}$.
        }
        \label{fig:example}
    \end{center}
\end{figure}

In Figure \ref{fig:example} \textbf{a)} an example of a DAG with two influential hidden variables is given.
In such a scenario, the MAG and PAG (Fig. \ref{fig:example} \textbf{b), c)}) are dense and the PAG contains many uninformative circle edgemarks.
For comparison, Figure \ref{fig:example} \textbf{d)} depicts our target object $\mathcal{C}_O$ which is sparse and contains more informative edge marks.

We will use the following standard notations. For an arbitrary matrix $M$,
$\norm{M}_1$ denotes the sum of its entries' magnitudes; $\norm{M}_\ast$
is the sum of its singular values; $\norm{M}_\infty$ is its largest entry
in magnitude; $\norm{M}_2$ is its largest singular value; $\norm{M}_F$ is the
Frobenius norm.
In addition, for a symmetric matrix $M$, $M \succ 0$ (resp. $M \succeq 0$) indicates that
$M$ is positive definite (resp. positive semi-definite).
We denote by $\text{degree}(M)$ the maximum number of
non-zero entries in any row or column of $M$. 
If $\mathcal{G}$ is a partially directed graph and $M$ is the adjacency matrix
of its skeleton, we define its degree as $\text{degree}(\mathcal{G}) := \text{degree}(M)$.

%% file: Figures/Example/example_DAG.tex
    \xymatrix{& *++[o][F.]{H_1} \ar@{.>}[ld] \ar@{.>}[dd] \ar@{.>}[rrd]& *++[o][F.]{H_2} \ar@{.>}[ldd] \ar@{.>}[rd] \ar@{.>}[rdd] \\
    X_1 \ar[rd] & & X_3 \ar[ld] & X_4 \ar[d]\\
        &X_2&&X_5
    }

%% file: Figures/Example/example_MAG.tex
    \xymatrix{
        X_1 \ar[rd] \ar@/^2pc/@{<.>}[rrr] \ar@{<.>}[rrrd] & & X_3 \ar[ld] & X_4 \ar[d]\\
        &X_2 \ar@/_1pc/@{<.>}[rr] \ar@{<.>}[rru]&&X_5 
    }

%% file: Figures/Example/example_PAG.tex
    \xymatrix{
        X_1 \ar@{*{\xyO}->}[rd] \ar@/^2pc/@{*{\xyO}.*{\xyO}}[rrr] \ar@{*{\xyO}.*{\xyO}}[rrrd] & & X_3 \ar@{*{\xyO}->}[ld] & X_4 \ar@{*{\xyO}-*{\xyO}}[d]\\
        &X_2 \ar@/_1pc/@{<.*{\xyO}}[rr] \ar@{<.*{\xyO}}[rru]&&X_5 
    }

%% file: Figures/Example/example_CPDAG.tex
    \xymatrix{
        X_1 \ar[rd] & & X_3 \ar[ld] & X_4 \ar@{*{\xyO}-*{\xyO}}[d]\\
        & X_2 && X_5 
    }

%% file: Sections/SuggestedEstimator.tex
\subsection{Suggested Estimators}

In this section, we discuss methods for estimating $K_O^{-1}$ and $\mathcal{C}_O$ under Settings 1 and 2. To this end, we first discuss the problem of estimating $K_O$. Recall that we denote the marginal covariance matrix of $X_O$ by $\Sigma$, and that its inverse $\Sigma^{-1}$ equals $K_O - L$.
Even in the absence of noise, inferring the components of $K_O - L$
 is a challenging problem because it is fundamentally misspecified: an infinity of pairs $(\hat{K}_O, \hat{L})$
satisfy the equation $K = \hat{K}_O - \hat{L}$ under the constraints $\hat{K}_O - \hat{L} \succ 0$, $\hat{L} \succeq 0$.

For $C^\ast$ an arbitrary matrix such that $C^\ast = A^\ast + B^\ast$, the problem of recovering
$A^\ast$ and $B^\ast$ from $C^\ast$ or an estimate of $C^\ast$ has been studied when $A$ is sparse and $B$ is dense and of low-rank \citep{CandesEtAl11, ChandrasekaranEtAl11}.
Loosely speaking, they showed that $(A^\ast, B^\ast)$ is with high probability equal to the solution of the convex problem,
\begin{equation} \label{noiseless_estimator}
  \argmin_{A,B}~ \gamma\norm{A}_1 + \norm{B}_\ast,~\text{~such that~}~C^* = A + B,
\end{equation}
provided $\gamma$ is chosen
within a suitable interval. The form taken by \eqref{noiseless_estimator} is motivated by the fact that the $\ell_1$ and
nuclear norms are convex relaxations for the $\ell_0$-norm and the rank respectively.
The penalties on $\norm{A}_1$ and $\norm{B}_\ast$ encourage the learning of a sparse $A$
and a low-rank $B$, while the tuning parameter $\gamma$ adjusts the relative
weight of these two penalties.
In the special case of multivariate Gaussian distributions, \cite{ChandrasekaranParriloWillsky12} showed that it is also possible to recover $K_O$ and $L$
when only samples from the marginal distribution of $X_O$ are available.
In this context, the assumption that $L$ is dense and low-rank means that there
must be relatively few hidden variables with an effect spread over most of the observed variables.
An estimate $(\hat{K}_O, \hat{L})$ of
$(K_O, L)$ is obtained as the minimiser of a function which couples the Gaussian log-likelihood:
with \eqref{noiseless_estimator}
\begin{equation}\label{lrps_estimator}
    (\hat{K}_O, \hat{L}) = \argmin_{(A,B) \in \mathbb{R}^{p\times p} \times \mathbb{R}^{p\times p}} -\ell(A - B; \hat{\Sigma}^{samp}_n) + \eta_n(\gamma ||A||_1 + ||B||_\ast)~\text{~such that}~A - B \succ 0,~B \succeq 0,
\end{equation}
where $\ell(K;\hat{\Sigma}^{samp}_n) = -\trace(K \hat{\Sigma}^{samp}_n) + \log \det K$ and $\eta_n,\gamma > 0$.
Here, the Gaussian log-likelihood makes it possible to learn an inverse covariance from the sample covariance
$\hat{\Sigma}^{samp}_n$, while the penalty plays the double role of regularising the
likelihood to prevent singularities (via $\eta_n$) and decomposing the estimated
precision matrix into its components.
The objective function in \eqref{lrps_estimator} is jointly convex in its parameters and can
be efficiently minimised even when $p$ is in the thousands \citep{Ma:2013}.
We call this estimator the ``low-rank plus sparse'' estimator (LRpS) and we write $LRpS(\eta_n, \gamma; \hat{\Sigma}_n)$ for the program which applies \eqref{lrps_estimator} to a positive semi-definite matrix $\hat{\Sigma}_n$, with tuning parameters $\eta_n, \gamma$
and outputs a pair of matrices $(\hat{K}_O, \hat{L})$.

When the random variables
are jointly Gaussian, zero partial correlation and conditional independence are equivalent.
This puts the edges of a Gaussian graphical
model and the non-zero entries of the precision matrix in a one-to-one correspondence \citep{Lauritzen96}.
This property is desirable but is not necessary for \eqref{lrps_estimator} to consistently estimate $K_O$ -- and therefore irrelevant to the problem at hand.
All that is required is a consistent estimator of $\Sigma$.
When the errors are sub-Gaussian, the sample covariance matrix is such an estimator \citep{Vershynin12}.
For heavy-tailed distributions, a modified Kendall correlation matrix
can be used \citep{Liu2012}. 


Provided the conditions for consistency of LRpS are met, an algorithm which assumes causal sufficiency can be readily applied to the estimated covariance matrix ${\hat{K}}_O^{-1}$ for estimating $\mathcal{C}_O$ \citep{Spirtes98, NandyHauserMaathuis16}. 
For structure learning, we suggest using the Greedy Equivalence Search (GES) algorithm which performs a greedy search to optimize an $\ell_0$-regularised log-likelihood score \citep{Chickering02}.
Let us write $GES(\lambda_n, \hat{A})$ for the program which applies GES to a covariance matrix $\hat{A}$ with tuning parameter $\lambda_n$ and outputs a CPDAG $\hat{\mathcal{C}}_O$.
The suggested estimator, called \emph{LRpS+GES} henceforth, can be summarised as in Algorithm \ref{alg1}.
We will show that it is consistent in some high-dimensional regimes when the data is generated 
according to Setting 1.
\begin{algorithm}[!h]
    \caption{Description of the LRpS+GES estimator}
    \label{alg1}
\begin{algorithmic}
\REQUIRE Sample covariance matrix $\hat{\Sigma}^{samp}_n$, tuning parameters: $\eta_n > 0,~\gamma > 0,~\lambda_n > 0$.
\ENSURE $\hat{\mathcal{C}}_O$, an estimate of the true CPDAG $\mathcal{C}_O$ of $\mathcal{G}_O$.
\STATE 1 - $(\hat{K}_O, \hat{L}) \leftarrow LRpS(\eta_n, \gamma; {\Sigma}^{samp}_n)$.
\STATE 2 - $\mathcal{C}_O \leftarrow GES(\lambda_n; \hat{K}_O^{-1})$.
\end{algorithmic}
\end{algorithm}
For Setting 2, we suggest an algorithm (Algorithm \ref{alg2}) which replaces the sample covariance matrix by a rank-based correlation matrix
and prove its high-dimensional consistency when the errors follow an elliptical distribution.
We call the resulting algorithm \emph{Kendall-LRpS+GES} (Algorithm \ref{alg2}).
\begin{algorithm}[!h]
    \caption{Description of the Kendall-LRpS+GES estimator}
    \label{alg2}
\begin{algorithmic}
\REQUIRE Sample Kendall correlation matrix $\hat{T}_n$, tuning parameters: $\eta_n > 0,~\gamma > 0,~\lambda_n > 0$.
\ENSURE $\hat{\mathcal{C}}_O$, an estimate of the true CPDAG $\mathcal{C}_O$ of $\mathcal{G}_O$.
\STATE 1 - $\hat{\Sigma}^{\tau}_n \leftarrow \sin\left(\frac{\pi}{2} \hat{T}_n \right)$, where the $\sin$ function
is applied elementwise.
\STATE 2 - $\hat{\Sigma}^{\tau+}_n \leftarrow \argmin_{S \in \mathcal{F}^p} \norm{S - \hat{\Sigma}^{\tau}_n}_F$,
where $\mathcal{F}^p$ is the space of correlation matrices of size $p$.
\STATE 3 - $(\hat{K}_O, \hat{L}) \leftarrow LRpS(\eta_n, \gamma; \hat{\Sigma}^{\tau+}_n)$.
\STATE 4 - $\hat{\mathcal{C}}_O \leftarrow GES(\lambda_n; \hat{K}_O^{-1})$.
\end{algorithmic}
\end{algorithm}

At a practical level, the fact there are three tuning parameters might be a legitimate concern.
We suggest first selecting the tuning parameters of $LRpS$ -- $\eta_n, \gamma$ -- using cross-validation or the (extended) BIC \citep{FoygelDrton10} and then choosing $\lambda_n$, so that there is no need to scout a 3-dimensional grid.
Moreover, we will see that both theoretical and empirical results support the idea that LRpS is not very sensitive to the value of $\gamma$: trying only a few values (\emph{i.e.} five or so) of this tuning parameter is enough for most applications --
more practical details are given later.
Finally, we note that the second step of Algorithm \ref{alg2} can be performed
 efficiently (see \cite{Qi06} and references therein).
We use the solver suggested in \cite{Qi06}\footnote{Available at \url{http://www.math.nus.edu.sg/~matsundf/} .}.

It might be a bit surprising that we can estimate $\mathcal{C}_O$ from an estimate of $K_O^{-1}$ regardless of the distribution of $\varepsilon$ in \eqref{eq:subG_SEM} or in \eqref{eq:trans_SEM}. However, as noted by \cite{Spirtes98, NandyHauserMaathuis16}, if $Z = (Z_1,\ldots, Z_p)$ is generated from a linear SEM with uncorrelated errors and $\mathcal{G}$ is a perfect map of the distribution of $Z$, then regardless of the distribution of the error variables
\[
    Z_i \indep Z_j | \{Z_r | r \in U \}~\Leftrightarrow~ Z_i \dsep_{\mathcal{G}_O} Z_j | \{Z_r | r \in U \}~\Leftrightarrow~\rho_{ij|U} = 0,
\]
where $i \neq j$, $U \subseteq \{1,\ldots,p\} \setminus \{i,j\}$ and $\rho_{ij|U}$ denotes the partial correlation between
$Z_i$ and $Z_j$ given $\{Z_r | r \in U \}$. Under Settings 1 and 2, we can draw the same conclusion by setting $Z = (X_O \mid X_H = x_H)$ or $Z = (f_O(X_O) \mid X_H = x_H)$ for all values of $x_H$ in the range of $X_H$. This enables us to learn $\mathcal{C}_O$ from partial correlations defined by the covariance (or correlation) matrix $K_O^{-1}$.


%% file: Sections/PreviousWork.tex
\subsection{Previous Work}

Over the past two decades, significant advances have been made on the problem of estimating DAGs from observational data.
This is a task which is known to be challenging, especially in the high-dimensional setting.
For example, the space of DAGs is non-convex and its size increases super-exponentially with the dimension of the problem \citep{Robinson1977}.
Structure learning algorithms fall into three main categories that we review here.
Since there are many approaches in each of these categories we refer the reader to \cite{Han2016}, \cite{Drton2017}, \cite{Heinze2018} for a more detailed overview and simulation studies.

\emph{Score based} approaches assign a score to each structure and aim to identify the one (or ones) that maximises a scoring function.
Usually, the scoring criterion measures the quality of a candidate structure based on the data.
Due to its theoretical properties and its performance on real and simulated datasets, we give special attention to the GES algorithm of \cite{Chickering02}.
GES is a greedy algorithm which searches for the CPDAG that maximises the $\ell_0$-penalised log-likelihood score over the space of CPDAGs.
It proceeds with a forward phase in which
single edge additions are carried out sequentially so as to yield the largest possible increase of the score criterion, until no addition can improve the score further.
The algorithm then starts with the output of the forward phase and uses best single edge deletions until the score can no longer be improved.
In spite of being a greedy algorithm, GES is consistent not only in the classical sense (``fixed $p$, increasing $n$'') \citep{Chickering02} but also in certain sparse high-dimensional regimes \citep{NandyHauserMaathuis16}.

\emph{Constraint based} algorithms learn graphical models by performing conditional independence tests.
The Peter Clark (PC) algorithm is a popular approach that falls in this category \citep{SpirtesEtAl00}.
Under suitable conditions, it is consistent for CPDAG recovery, even in the high-dimensional regime \citep{KalischBuehlmann07a,HarrisDrton13,ColomboMaathuis14}.
When there are hidden variables and/or selection bias, the counterpart of the PC-algorithm is the Fast Causal Inference (FCI) algorithm whose output is a partial ancestral graph \citep{SpirtesEtAl00, RichardsonSpirtes02}.
While consistent in sparse high-dimensional settings \citep{ColomboEtAl12}, FCI is not fast enough to be applied to large graphs.
This limitation prompted the development of methods such as the \emph{Really Fast Causal Inference} (RFCI) algorithm and FCI+ \citep{ColomboEtAl12,ClaassenMooijHeskes13}.
A strength of FCI-type algorithms is that the hidden structure can be arbitrarily complicated, since no assumptions are made about selection bias and hidden variables.

\emph{Hybrid algorithms} combine constraint-based and score-based methods.
For example, the Max-Min Hill-Climbing (MMHC) algorithm first learns the skeleton using a local discovery algorithm and then orients the edges via a greedy hill-climbing procedure \citep{TsamardinosEtAl06}.
The NSDIST approach suggested in \cite{Han2016} also outputs a DAG in two-stages.
In the first stage, the adaptive-lasso \citep{Zou2006} is used to perform neighbourhood selection.
For the second stage, \citet{Han2016} suggest a novel greedy algorithm which searches the space of DAGs whose neighbourhoods agree with the output of the first stage.
Finally, the adaptively restricted GES of \cite{NandyHauserMaathuis16} is a hybrid approach which modifies the forward phase of GES by
adaptively restricting the search space. They show that this approach remains consistent in some sparse high-dimensional regimes, and is faster than GES.

In summary, constraint based methods come with theoretical results 
assuming none or arbitrarily many hidden variables. 
This is different from the set-up assumed here in that we wish to 
consider an intermediate setting where there are few confounders 
with widespread effects.
As for score-based and hybrid methods, most work assumes that there 
are no hidden confounders.

Finally, the type of confounding we consider in this paper in ubiquitous is genomics applications, which is
why the problem of estimating and removing this kind of unwanted variation has been well studied \citep{Leek2007, Speed2013,Mostafavi2013}.
The work of \cite{ChandrasekaranParriloWillsky12} on which we build is also applicable to 
this problem and has been available for a few years. 
However, to the best of our knowledge, it has never been applied to
causal structure learning.
In that respect, the approach of \cite{Silva2006} is closer to what 
is suggested here in that they aim at estimating a linear DAG in the presence of latent variables under some assumptions about the relationship between observed and unobserved variables. 
A simple solution to our problem consists in estimating the first few principal components of the data and to regress them out before conducting any analysis.
More sophisticated, general purpose algorithms have also been developed. PEER, for example, is a Bayesian approach which aims at inferring
``hidden determinants and their effects from gene expression profiles by using factor analysis methods'' \citep{Stegle2012}.
It was recently used by the GTEX consortium in order to remove confounding from their datasets \citep{Aguet2016}.
In what follows, our work will be compared to both the principal component analysis and the PEER approaches.

%% file: Sections/Consistency.tex
\label{section:consistency}

The high-dimensional behaviour of the Low-Rank plus Sparse decomposition (LRpS henceforth) and the
 GES algorithm has been well studied  
 \citep{ChandrasekaranParriloWillsky12, NandyHauserMaathuis16}.
We rely on this body of work to derive the high-dimensional consistency of LRpS and LRpS+GES
for sub-Gaussian random vectors and transelliptical distributions.

We consider an asymptotic scenario where
both the dimension of the problem and the sample size are allowed to grow simultaneously,
meaning that the number of observed variables $p$ and the number of hidden variables $h$ are now functions of $n$. We write $p_n$ and $h_n$ to make this dependence explicit.
Likewise, we write $X_{On} \in \mathbb{R}^{p_n}$
for the random vector being modelled.
We also write $\tK,~\tL$ and $\tC$ to make it clear that
the nominal parameters are indexed by $n$. The same holds for the estimates obtained from Algorithms \ref{alg1} and \ref{alg2}
$(\eK, \eL, \eC)$.
We let $\tR$ be the true partial correlation computed from $K_{nO}^{-1}$ between the $i$-th and the $j$-th variable given the variables in a set of indices $U$,
for $i,j \in \{1,\ldots,p_n\}$ and $U \subseteq \{1,\ldots,p_n\} \setminus \{i,j\}$. These partial correlations correspond to partial correlations in a sub-Gaussian (Setting 1) or an elliptical distribution which has a covariance or a correlation matrix equals $K_{nO}^{-1}$.
The sample partial correlation $\eR$ is defined similarly based on an estimated sample covariance/correlation matrix $\hat{\Sigma}_n$. We choose $\hat{\Sigma}_n$ to be the sample covariance matrix $\Sigma_n^{samp}$ for Setting 1, and choose $\hat{\Sigma}_n$ to be $\hat{\Sigma}_n^{\tau} : = \sin(\frac{\pi}{2} \hat{T}_n)$ for Setting 2 where $\hat{T}_n$ denotes the sample Kendall correlation matrix.

We prove the following results in Appendix A.
The proof first proceeds by establishing the consistency of LRpS 
in Settings 1 and 2. We provide convergence rates for the recovery of $K_{nO}$ in terms of the max-norm ($\norm{\cdot}_\infty$).
Building on these preliminary results, we derive the convergence rate for $K_{nO}$ in spectral norm and, in turn, the convergence 
rate of ${K_{nO}}^{-1}$ in spectral norm. We then build on the work
of \citet{NandyHauserMaathuis16} to conclude. 

\begin{theorem}
\label{theo1}
  Assume that the data is generated according to Setting 1: $X_{n}$ is jointly sub-Gaussian 
  and $\mathcal{G}_{nO}$ is a perfect map for the distribution of $X_{nO}$ conditional on $X_{nH}$,
  as described by Equation \eqref{eq:subG_SEM}.

  Assume (A1), (A2), (A6) and (A6') given below and 
  let $\hat{K}_{nO}$ and $\hat{\mathcal{C}}_{nO}$ be as in Algorithm \ref{alg1}.
  Then there exists a sequence 
  $\eta_n$ such that $\norm{K_{nO} - \hat{K}_{nO}}_\infty = O_P \left(\sqrt{\frac{p_n}{n}}\right)$, for a suitable choice of $\gamma$.

  Assume further that (A3) - (A5) hold. 
  Then there exists a sequence $\lambda_n$ such that
$\mathbb{P}\left(\eC = \tC \right) \xrightarrow[n\to\infty]{}1$.\\
\end{theorem}

\begin{theorem}
  \label{theo2}
    Assume that the data is generated according to Setting 2: $X_{n}$ is jointly transelliptical 
    and $\mathcal{G}_{nO}$ is a perfect map for the distribution of $X_{nO}$ conditional on $X_{nH}$,
    as described by Equation \eqref{eq:trans_SEM}.
  
    Assume (A1), (A2) and (A6) given below and 
    let $\hat{K}_{nO}$ and $\hat{\mathcal{C}}_{nO}$ be as in Algorithm \ref{alg2}.
    Then there exists a sequence 
    $\eta_n$ such that $\norm{K_{nO} - \hat{K}_{nO}}_\infty = O_P \left(\sqrt{\frac{p_n \log p_n}{n}}\right)$, for a suitable choice of $\gamma$.
  
    Assume further that (A3) - (A5) hold. 
    Then there exists a sequence $\lambda_n$ such that
  $\mathbb{P}\left(\eC = \tC \right) \xrightarrow[n\to\infty]{}1$.\\
  \end{theorem}

Assumptions (A1) - (A6) and (A6') are as follows:
\begin{description}
    \item[(A1)] (Consistency of LRpS) The conditions for the algebraic consistency of LRpS are satisfied
    (see Theorem 4.1 of \cite{ChandrasekaranParriloWillsky12} and conditions (LRpS1,2) in Appendix A). One implication is that
    one requires \emph{at least} $n \geq 2 p_n$ (Th. \ref{theo1}) or $n \geq p_n \log p_n$  (Th. \ref{theo2}).
    \item[(A2)] (Scaling Regime) $p_n = \mathcal{O}(n^{1-a})$, for some $0 < a < 1$.
    \item[(A3)] (Sparsity condition) Let $q_n = \text{degree}(\tC)$ and
    $q_n' = \text{degree}(\tK)$. Then $q_n \leq q_n'$.
    We assume that $q_n' = \mathcal{O}(\log(n)^b)$, for some $0 \leq b \leq \infty$.
    \item[(A4)] (Bounds on the growth of the oracle versions) The maximum degree in the output of the forward phase
    of every $\delta_n$-optimal oracle version of GES is
    bounded by $K_nq_n =  \mathcal{O}(n^{1-f})$, for some
    sequence $\delta_n^{-1} = \mathcal{O}(n^{d_1})$ such that $0 < f \leq 1$ and $0 \leq 2 d_1 < a$, and where $q_n$ is given by (A3) and $a$ is given by (A2).
    \item[(A5)] (Bounds on partial correlations) The partial correlations $\tR$ computed from $K_{nO}^{-1}$ satisfy
    the following upper and lower bound for all $n$ and $U \subseteq \{1,\ldots,p_n\} \setminus \{i,j\}$ such that $|U| \leq K_n q_n$:
   \begin{align*}
    \mathop{\sup}_{i\neq j, U} |\tR| \leq M < 1,  ~~ \text{and} ~~  \mathop{\inf}_{i, j, U} \{|\tR| : \tR \neq 0 \} \geq c_n,\vspace{-0.05in}
   \end{align*}
   with $c_n^{-1} = \mathcal{O}(n^{d_2})$ for some $0 \leq 2 d_2 < a$ where $a$ is as in (A2).

    \item[(A6)] $\norm{{\tK}^{-1}}_2 < C_4$ and $\norm{\tK}_\infty < C_5$, for some $C_4, C_5 \geq 0$.
    \item[(A6')] The sub-Gaussian norm of $X_{nO}$ is bounded above 
    by an absolute constant.
\end{description}

In the previous section, it was mentioned that the LRpS estimator is consistent when $\tK$ is sparse and $\tL$ is dense and low-rank.
Assumption (A1) contains more precise requirements for the problem to be identified.
One of the conditions for identifiability is expressed as
$\xi(\tL) \mu(\tK) \leq  \frac{1}{6}C^2$, for some constant ${C}$ which depends on the Fisher information matrix.
Here, $\xi(\tL)$ is a property of $\tL$ such that a small value of $\xi(\tL)$ guarantees that no single hidden variable will have
    an effect on only a small number of the observed variables.
    It is related to the concept of \emph{incoherence}, which is easily calculated and satisfies
    $inc(M) \leq \xi(M) \leq 2inc(M)$, for any matrix $M$ \citep{CandesEtAl11, ChandrasekaranParriloWillsky12}.
      On the other hand, $\mu(\tK)$ quantifies the diffusivity of $\tK$'s spectrum.
   Matrices that have a small $\mu$ have few non-zero entries per row/column.
   Thus, (A1) entails that there must be few hidden variables acting on many observed ones and that $\tK$ must
have sparse rows/columns.
  Assumption (A1) also requires that the tuning parameter $\gamma$ be chosen such that $\gamma = \frac{C}{2\mu(\tK)}$, which implies that 
  the sample size must satisfy $n \geq A \mu^4(K_{nO}) p_n$ (Th. \ref{theo1}) or $n \geq A \mu^4(K_{nO}) p_n \log p_n$ (Th. \ref{theo2}), 
  for some absolute constant $A$ (see Appendix A). Since $\mu(K_{nO})$
  is expected to increase with the degree of $K_{nO}$, this shows
  that the requirement on the minimum sample size increases typically increases with the number of edges of $K_{nO}$.

An important feature of Theorem \ref{theo1} is that the degrees of the true CPDAG
and $\tK$ are allowed to grow logarithmically with the sample size $n$.
When coupled with (A1), this assumption on the growth rate of $q_n'$ imposes restrictions on the number of hidden variables $h_n$, albeit not explicitly.
Indeed, it can be shown that for the condition $\xi(\tL) \mu(\tK) \leq  \frac{1}{6}C^2$
to hold with high probability, $h_n$ has to be of the form $h_n = \mathcal{O}(\frac{p_n}{\log (p_n)^{2d}})$ (under some assumptions about the distribution from which $\tL$
is sampled) \citep{ChandrasekaranParriloWillsky12}.
Thus, the degree of $\tC$ and the number of hidden variables are allowed to grow simultaneously 
with the sample size, and in that regime, $n \sim p_n \log p_n$ samples are required 
for consistent estimation (see Appendix A).
A similar conclusion can be drawn for Theorem \ref{theo2}.

The rate of $\sqrt{\frac{p_n \log p_n}{n}}$ in Theorem \ref{theo2} is due to the 
recent results established by \cite{Wegkamp2016} and \cite{han2017} for the convergence in spectral 
norm of the modified Kendall correlation matrix.
As mentioned above, 
$p_n \log p_n$ samples are necessary for the consistent estimation of a latent Gaussian graphical
model.
Therefore, the Kendall-LRpS+GES estimator -- whose rate is inflated only by a factor of 
$\sqrt{\log p_n}$ -- is consistent under conditions that are almost identical to LRpS+GES 
since $n \sim  p_n \log p_n$ is already required in the sub-Gaussian setting.
Thus, the scaling regime of (A2) is strong enough to guarantee the consistency of 
both algorithms.

Finally, note that (A4) follows from (A1) with f = a, since the maximum degree in the output of the forward phase
    of every $\delta_n$-optimal oracle version of GES is always bounded by $p_n - 1 = \mathcal{O}(n^{1 - a})$. However, we keep (A4) as a separate assumption in order to facilitate a direct comparison between our assumptions and the corresponding assumptions of \cite{NandyHauserMaathuis16}.

%% file: Sections/Simulations1.tex
\subsection{CPDAG Structure Recovery}
\label{section:CPDAGRecovery}
Throughout, we generate DAGs with $p + h$ nodes, and set $p=50$ -- a value which does not depend on the sample size\footnote{The code for our simulations and applications is made available with this paper.}. 
In particular, our data is generated according to linear structural equation models of the form 
\[ X \leftarrow \begin{pmatrix} B_O & B_{OH} \\ 0 & B_H \end{pmatrix}X + \epsilon, \]
where $B_O \in \mathbb{R}^{p \times p}, B_{OH} \in \mathbb{R}^{p \times h}$ and 
$B_H \in \mathbb{R}^{h \times h}$ are matrices encoding the structure and effect 
sizes of the DAGs and $\epsilon \sim \mathcal{N}(0, \Lambda)$ \citep{bollen1989}. 
Furthermore, $B_O$ and $B_H$ are strictly upper-triangular matrices and 
$\Lambda \in \mathbb{R}^{(p +h) \times (p + h)}$ is a diagonal matrix.
The DAGs over the observed variables ($\mathcal{G}_O$) are random DAGs with an expected sparsity of 
$5\%$, which corresponds to an average degree of about 2.5 and an average maximum degree of about 6.3. 
The $h$ hidden variables remain independent, but each of them has directed edges towards a random 
$\zeta \%$ of the observed variables.
All edge weights -- \emph{i.e} the non-zero entries of the $B_{\cdot}$ matrices -- 
are drawn uniformly at random from $[-1, 1]$. Residual variances -- \emph{i.e.} the diagonal 
entries of $\Lambda$ -- follow a uniform distribution over $[0, 1]$.

In this section, we compare methods based on the precision-recall curves obtained by varying the tuning parameter for the last stage of the structure learning methods. The tuning parameters of the first stages (when applicable) are selected as described below.
The following methods are applied to the data:
\begin{description}
    \item \emph{GES} \citep{Chickering02}: implemented in the \texttt{pcalg} package \citep{KalischEtAl12}. 
    \item \emph{NSDIST} \citep{Han2016}: we used the code made available with the original article. For the tuning parameters of the first stage (called $\lambda_0$ and $\gamma$ in \cite{Han2016}), we used the values suggested in \cite{Han2016}. 
    \item \emph{PCA*+GES}: the top $k$ principal components are first estimated from the data matrix and regressed out. GES
    is then applied to the residuals. The number of principal components is chosen with \emph{perfect knowledge} (hence the * in the name) so as to maximise
    the average precision. 
    \item \emph{PEER*+GES} \citep{Stegle2012}: similar to PCA*+GES, the first stage is replaced with PEER. Here again, 
    the number of latent factors is selected so as to maximise average precision, hence the * in the name.
    \item \emph{LRpS+GES}: the suggested approach described in Algorithm \ref{alg1}. The tuning parameters $\eta_n, \gamma$ for LRpS are chosen by
    cross-validation with $\gamma \in \{0.05, 0.1, 0.15, 0.2, 0.3, 0.5, 0.7\}$. 
\end{description}

In our first set of simulations, we investigate the effect of the sample size $n$ and the number of hidden variables $h$ on CPDAG recovery. 
We set $n \in \{50, 200, 2000\}$ and $h \in \{0, 5, 10\}$, but fix $\zeta$ to 70. 
For each of the nine possible $(n, h)$ pairs, we generate 50 distinct DAGs and draw $n$ samples from each of them, for a total of 450 datasets. 
This is a setting which is favourable to our
approach since the hidden variables impact a large fraction (70\%) of the observed ones.

\begin{figure}
    \centering
    \begin{center}
     \begin{subfigure}[t]{0.03\textwidth}
    \textbf{a)}
    \end{subfigure}
    \begin{subfigure}[t]{0.8\textwidth}
        \centering
        \adjustbox{valign=t}{\includegraphics[width=\textwidth]{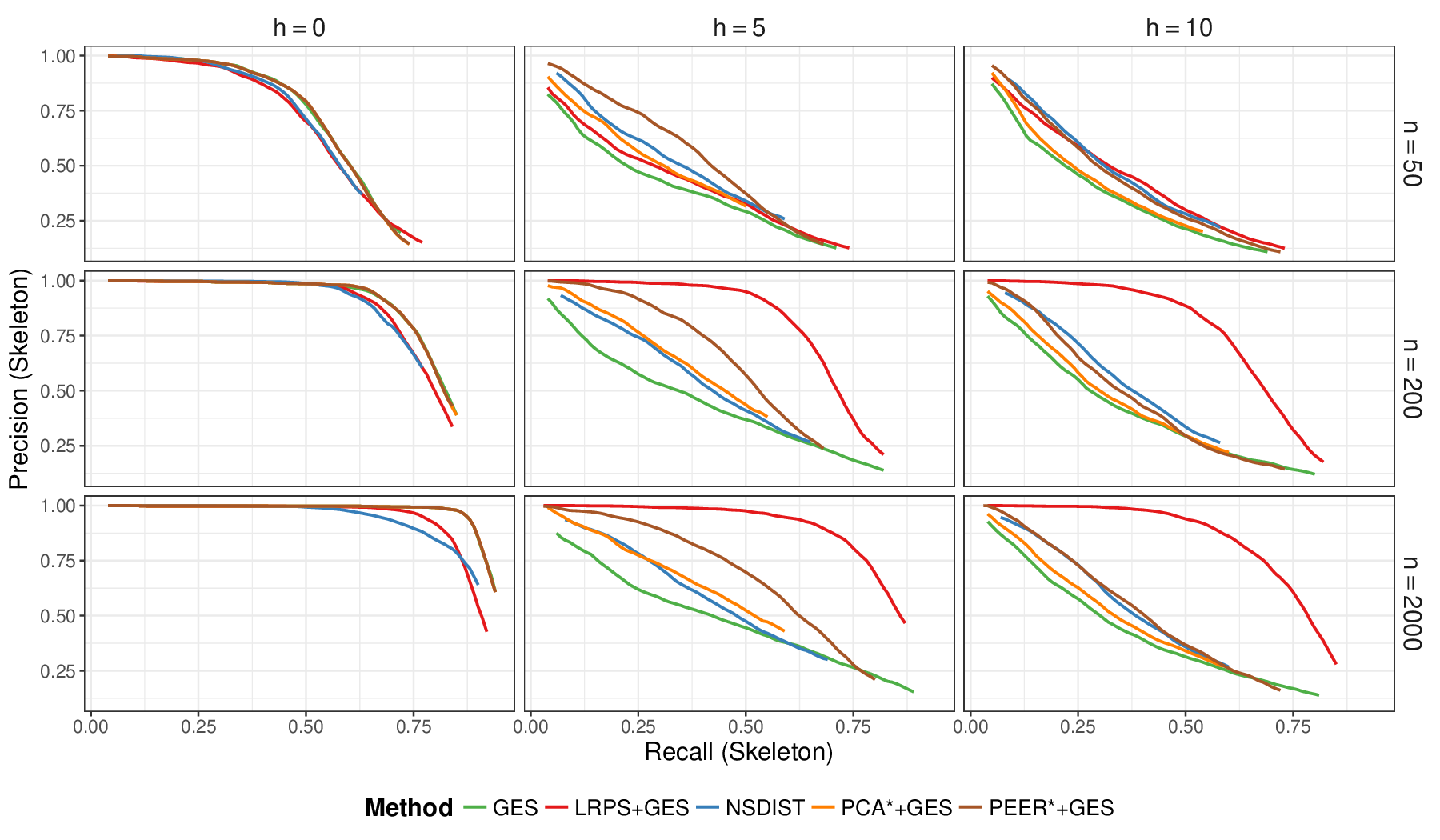}}
        \caption{}
    \end{subfigure}\\
         \begin{subfigure}[t]{0.03\textwidth}
    \textbf{b)}
    \end{subfigure}
    \begin{subfigure}[t]{0.78\textwidth}
        \centering
        \adjustbox{valign=t}{\includegraphics[width=\textwidth]{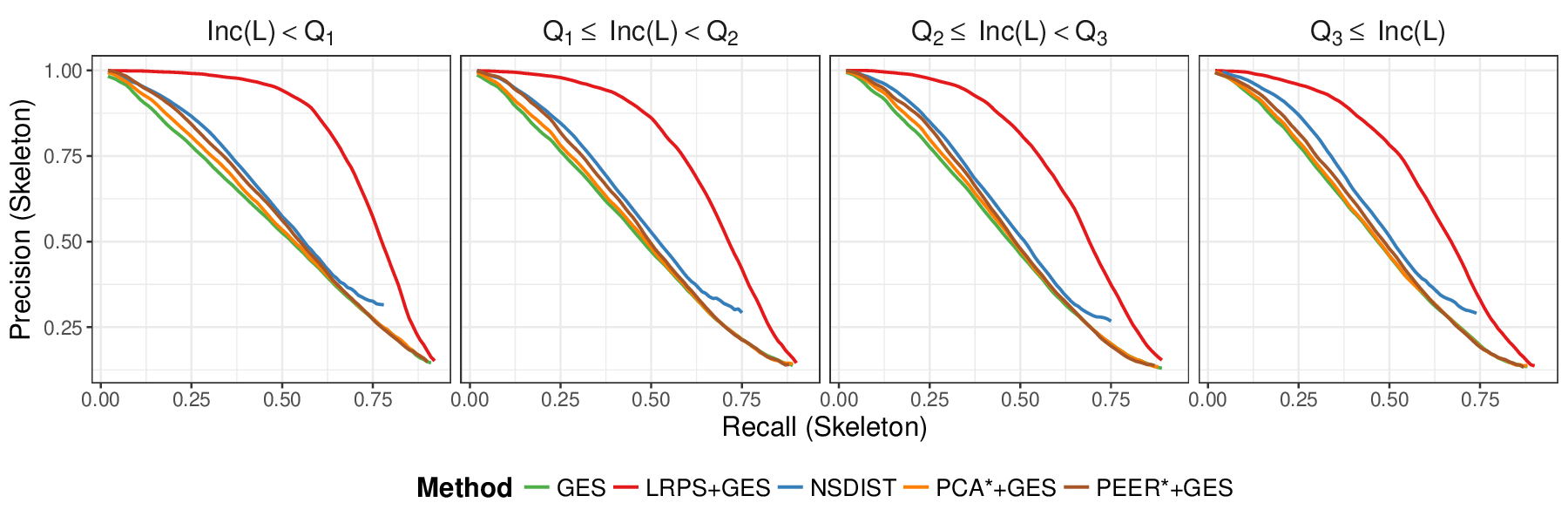}}
        \caption{}
    \end{subfigure} 
    \caption{Average precisions at fixed recalls of $\{0.01, 0.02, \ldots, 1\}$ for skeleton recovery. 
    There are $p=50$ observed variables. 
    \textbf{a)} Effect of the number of hidden variables $(h)$ and sample size $(n)$, when $\zeta = 70$ and each of the 9 designs is repeated 50 times.
    \textbf{b)} Effect of the incoherence of the latent structure $inc(L)$. The 500 random datasets
    are binned according to the quartiles of $inc(L)$'s distribution  ($Q_1,\ldots,Q_4$).
    }
    \label{FigureSim11}
    \end{center}
\end{figure}

In Figure \ref{FigureSim11} \textbf{a)} we assess the performances of the methods in terms of skeleton recovery by plotting average precision/recall curves.
Precision is calculated as the fraction of correct edges among the retrieved edges; recall is computed as the number of correctly retrieved edges 
divided by the total number of edges in the true CPDAG. 
Since each of the 9 designs is repeated 50 times, we report average precisions at fixed recalls of $\{0.01, 0.02, \ldots, 1\}$.
In the appendix, similar curves are plotted for directed edges. 
When there is no confounding (leftmost column), all methods are known to be consistent for skeleton recovery and offer comparable performances.
As soon as $h > 0$, GES is outperformed. 
Unsurprisingly, when $n$ and $p$ are of the same order of magnitude, PCA does not perform as 
well as a Bayesian approach like PEER. Overall, none of the methods offer good performances 
when $n = 50$ and there is confounding, as suggested by our theoretical results. 
When it comes to skeleton recovery, LRpS+GES is always at least as good as the other methods. 
When there is confounding, it is significantly better
because it is the only method which explicitly models hidden variables.
This is true even though the tuning parameters $\eta_n$ and $\gamma$ were chosen with cross-validation. 
We also see that when $h > 0$, LRpS+GES is the only method whose performance improves with 
increasing sample size.
It is, however, not consistent because the distribution of parameter values chosen in these simulations
is in clear violation of our assumptions, in particular the smallest eigenvalue of $L$ is too small
for this noise level.

In our second set of simulations, we draw from a more diverse set of hidden structures.
We set $n = 500$ but draw $h$ and $\zeta$ uniformly at random from $[5,30]$ and $[15, 70]$ respectively.
We generate 500 datasets according to this scheme.
In order to quantify the departure of $L$ from our assumptions we compute $inc(L)$, the incoherence of
 $L$, for each of the 500 datasets (the distribution of $inc(L)$ along with figures showing the 
 effect of $h$ are shown in the appendix).
In this second scenario, many datasets explicitly violate our assumptions since there are 
many hidden variables acting in a sparse fashion.

In Figure \ref{FigureSim11} \textbf{b)}, we plot average precision/recall curves for this second 
simulation design. 
The datasets are divided into four bins based on the
quartiles of $inc(L)$'s distribution (noted $Q_1, \ldots, Q_4$),
so that the leftmost panel corresponds to the 125 datasets for which it is easiest to estimate
$L$.
Doing so indicates how our approach is expected to behave in the most adverse scenarios. 
As can be seen from this figure, LRpS+GES outperforms other approaches in terms of skeleton recovery.

%% file: Sections/Simulations2.tex
\subsection{Total Causal Effect Estimation}

Under our assumptions
is the \emph{causal} DAG so that the Markov equivalence class encoded 
 by $\mathcal{C}_O$ contains the causal DAG.
 For any given pair of distinct nodes $(X_i, X_j)$, we can therefore estimate the total causal effect of $X_i$ on $X_j$ for all DAGs 
 in the Markov equivalence class. 
 Since one of these DAGs is the true causal DAG, this yields a list of possible total causal effects which includes the true total causal effect.
 The IDA approach makes it possible to generate such lists efficiently without 
 enumerating all DAGs in the Markov equivalence class \citep{MaathuisKalischBuehlmann09,Maathuis2010}. 
 The original IDA method described in \citet{MaathuisKalischBuehlmann09} 
 uses the PC algorithm in order to first estimate a CPDAG, and then computes sets of possible total causal effects 
 using the sample covariance matrix and the output of the first stage.
However, it is possible to replace this first step by any other algorithm which estimates a CPDAG. Likewise, 
any estimator of $\Sigma$ can replace $\hat{\Sigma}_n^{samp}$.

Since LRPS+GES outputs a CPDAG, we can assess its ability to estimate total causal effects
 by using it in the first stage of IDA. 
 Thus, lists of possible causal effects are generated using 
the estimated CPDAG $\hat{\mathcal{C}}_O$ and the covariance matrix $\hat{K}_O^{-1}$.
We denote this method by (LRPS+GES),IDA.
 For all pairs $(i,j) \in \{1,\ldots,p\}^2$, $i \neq j$,
we compute the set $S_{ij}$ of possible total causal effects of $X_i$ on $X_j$. 
Pairs of variables $(X_i,X_j)$ are then ranked according to $\min(\{|s|: s \in S_{ij} \})$. 
This ranking is compared to the true total causal effects using the precision and recall metrics, \emph{e.g.} ``precision at rank $k$''
would be the number of pairs $(X_i, X_j)$ that are in the top $k$ pairs 
and have a non-zero total causal effect in the true DAG, divided by $k$.

In this section, we select a single DAG, PAG or CPDAG along the regularisation paths in order to 
apply IDA or LV-IDA. Thus, we pick a value of the tuning parameters for both the first and second stages.
This is in contrast with the previous section where only the tuning parameters of the first stages (when applicable) 
were selected, while we reported precision-recall curves for the whole regularisation paths of the second stages.
We consider the following methods, where the first stage tuning parameters are selected as before (when applicable):
\begin{description}
    \item \emph{GES,IDA}: the CPDAG is estimated using GES. The tuning parameter $\lambda_n$ is chosen with the BIC. 
    IDA is applied with the resulting CPDAG and
    the sample covariance matrix $\hat{\Sigma}_n^{samp}$. 
    \item \emph{NSDIST,IDA}: the DAG is estimated using NSDIST and converted to a CPDAG.
    The tuning parameter for the second stage ($\lambda$, with the notations of \cite{Han2016}) is chosen using the BIC.
    IDA is applied with the resulting CPDAG and
    the sample covariance matrix $\hat{\Sigma}_n^{samp}$. 
    \item \emph{(PCA*+GES*),IDA}: the top $k$ principal components are first estimated from the data and regressed out. The CPDAG is estimated using GES on the residuals.
    The tuning parameter $\lambda_n$ is chosen with perfect knowledge so as to maximise
    the average precision (in terms of causal effect recovery). 
    IDA is applied with the resulting CPDAG and
    the covariance matrix of the residuals. 
    \item \emph{(LRpS+GES),IDA}: the CPDAG is estimated using LRpS+GES.
    The tuning parameter of the second stage $\lambda_n$ is chosen with the BIC. 
    IDA is applied with the resulting CPDAG and
    the covariance matrix ${\hat{K}_O}^{-1}$.
    \item \emph{RFCI,LV-IDA} \citep{ColomboEtAl12, Malinsky2017}: the PAG is estimated with RFCI.
    The significance level $\alpha$ for RFCI is given by $\alpha = \frac{0.5}{\sqrt{n}}$\footnote{In a number of cases, the LV-IDA
    algorithm, when applied to a single dataset, was still running after a few days of computation. 
    Given that we simulated data from hundreds of datasets, we could not experiment with many values of $\alpha$.}.
    LV-IDA is applied to the resulting PAG and the sample covariance matrix $\hat{\Sigma}_n^{samp}$.
    Whenever LV-IDA outputs an NA, the corresponding pair is not counted, \emph{i.e.} it is neither a true positive nor false positive.
    \item \emph{RANDOM,IDA}: one hundred random DAGs are generated from the same model as was used in the simulation. 
    Total causal effects are then estimated based on the resulting
    CPDAG and the sample covariance matrix $\hat{\Sigma}_n^{samp}$.
    We report the interval spanned by the 2.5-97.5 percentiles of the distribution of precisions at fixed recalls.
    \item \emph{EMPTY, IDA}: Causal effects are computed without adjustment, 
    which is equivalent to applying the \texttt{ida} function of the \texttt{pcalg} package to an empty graph and the 
    sample covariance matrix.
\end{description}
With respect to total causal effect estimation, we found (PEER*+GES*),IDA and (PCA* + GES*),IDA to be nearly undistinguishable, which is why PEER
is not reported here.
Moreover, note that since we are reporting results for the GES,IDA approach, we are not considering the ``PC,IDA'' method. 
Indeed, GES has been shown to have good finite sample performance, and recent high-dimensional consistency guarantees 
have been given in \citet{NandyHauserMaathuis16}.

\begin{figure}
    \centering
    \begin{center}
     \begin{subfigure}[t]{0.03\textwidth}
    \textbf{a)}
    \end{subfigure}
    \begin{subfigure}[t]{0.8\textwidth}
        \centering
        \adjustbox{valign=t}{\includegraphics[width=\textwidth]{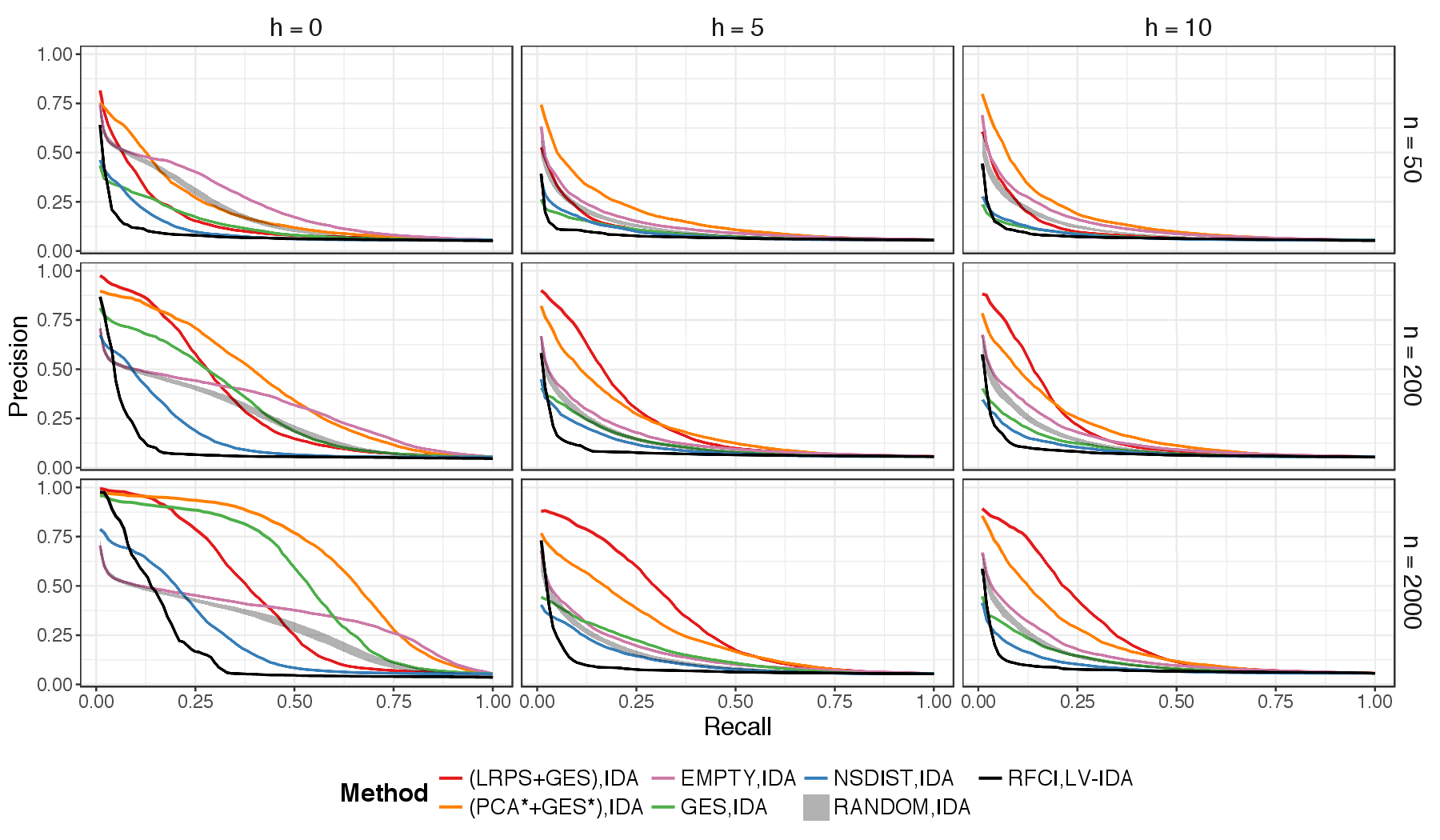}}
        \caption{}
    \end{subfigure}\\
         \begin{subfigure}[t]{0.03\textwidth}
    \textbf{b)}
    \end{subfigure}
    \begin{subfigure}[t]{0.78\textwidth}
        \centering
        \adjustbox{valign=t}{\includegraphics[width=\textwidth]{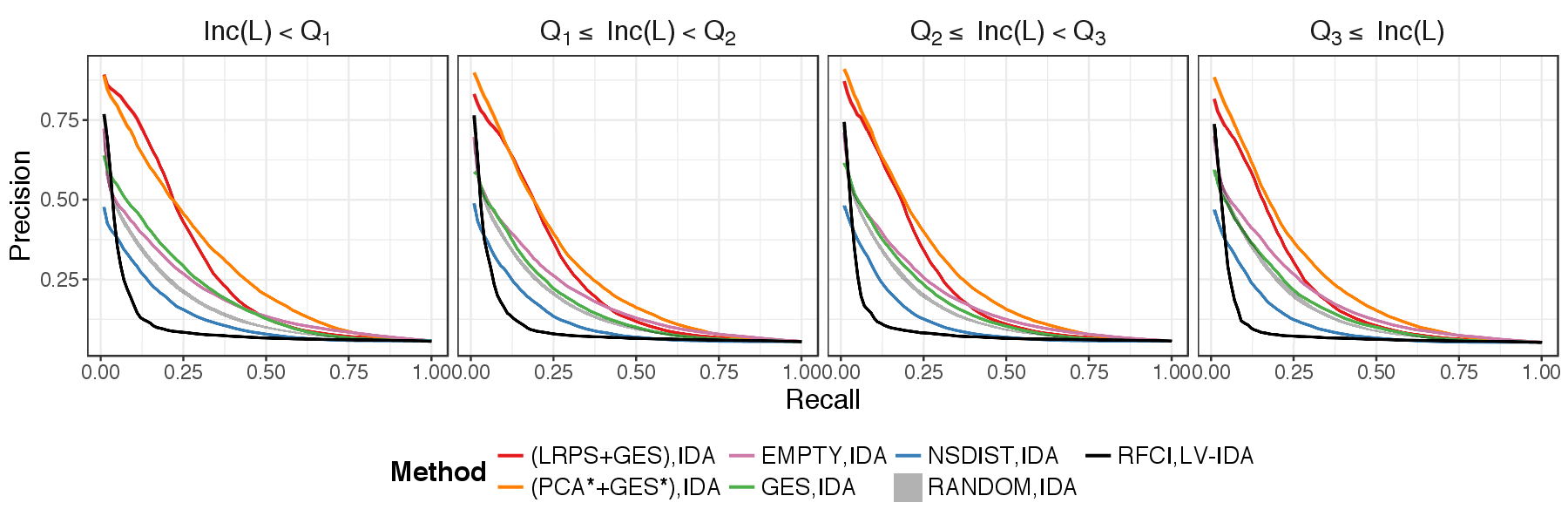}}
        \caption{}
    \end{subfigure} 
    \caption{Average precisions at fixed recalls of $\{0.01, 0.02, \ldots, 1\}$ for total causal effect recovery. 
    \textbf{a)} Effect of the number of hidden variables $(h)$ and sample size $(n)$ when $\zeta = 70$ and each of the 9 designs is repeated 50 times.
    \textbf{b)} Effect of the incoherence of the latent structure $inc(L)$. The 500 random datasets
    are binned according to the quartiles of $inc(L)$'s distribution  ($Q_1, \ldots, Q_4$).
    }
    \label{FigureSim2}
    \end{center}
\end{figure}

We consider the same simulation designs as in the previous section. 
Figure \ref{FigureSim2} \textbf{a)} displays the results obtained in the first setting, the one 
where hidden variables are influential. 
Unlike in Figure \ref{FigureSim11}, the orientation of the directed edges matters. 
When the sample size is relatively small $(n=50)$, 
there appear to be little to gain from using (CP)DAG estimation methods -- the EMPTY approach is competitive.
As soon as the sample size increases and $h > 0$, there is a clear benefit in using LRpS+GES. 
When $h=0$, LRpS+GES is outperformed but its performances
remain comparable to those of GES. 
Since it is designed to handle hidden variables, the behaviour of RFCI,LV-IDA might come as a surprise. 
First, we see that when there is no confounding, RFCI,LV-IDA is capable of achieving a high precision. 
This is consistent with previous findings indicating
that LV-IDA is conservative but capable of recovering a small but high-quality set of total 
causal effects \citep{Malinsky2017}. 
When $h>0$, the set of models we simulate from is particularly challenging for methods 
relying on MAGs since nearly all pairs of observed variables are confounded. 
It is therefore not surprising to see RFCI,LV-IDA being outperformed.

As can be seen from Figure \ref{FigureSim2} \textbf{b)}, LRpS+GES
performs at least as well other approaches and, in most cases, it performs better. 
As pointed out above, it is in the most challenging scenarios, when
confounders act in a sparse fashion (rightmost panel), that RFCI,LV-IDA is the most useful. It is very conservative but is capable of achieving the highest precision.

Finally, we recall that for the (PCA*+GES*),IDA method, both the number of 
principal components to regress out and the tuning parameter for GES $\lambda_n$ are chosen
so as to maximise the area under the precision/recall curves.
This provides a benchmark for the method, but such performances could not be achieved on 
a real dataset.
This explains the discrepancy between (PCA*+GES*),IDA and GES,IDA when $h=0$: GES,IDA 
selects $\lambda_n$ using the BIC.
It also puts into perspective the performances of (LRpS+GES),IDA and RFCI,LV-IDA, which
are sometimes far better than the other approaches 
in spite of selecting the tuning parameters from the data only.

%% file: Sections/Simulations3.tex
\subsection{Hubs, Robustness to Outliers and Non-Linearities}

In our simulations, we considered situations when the assumption on $L$ does not
hold, \emph{i.e.} when the hidden variables 
are not impacting a large fraction of the observed variables, but act in a sparse 
fashion instead.
Additionally, one can wonder what happens when the conditions on $K_O$ are not met, 
\emph{i.e.} the DAG over observed variables is not so sparse
and it has a high degree.
In the supplementary materials, we simulate random graphs from the Barabasi model and 
report results showing to what extent our approach is affected
by such graphs with hubs.
A summary of our findings is that LRpS+GES is indeed outperformed by 
GES when there are hubs with a high degree and no hidden variables.
When the hubs are of moderate size and there are 
hidden confounders, LRpS+GES 
remains superior to PEER and clearly outperforms GES.  
Finally, when hubs have a high degree and there are latent variables, the performance of all methods is degraded, but LRpS+GES is less affected than its counterparts.

We also looked at the performances of the Kendall-LRpS+GES estimator described in Algorithm \ref{alg2}
by simulating data contaminated with samples drawn from a Cauchy distribution (a violation
of our condition on $\mathbb{E} \xi$) and marginally 
transformed by strictly increasing functions ($x^3$), or non-monotonic functions ($x^2$) -- another
violation of our assumptions.
We found Kendall-LRPS+GES to be especially robust to outliers, even in the presence of hidden variables.
When variables are marginally transformed with a non monotonic function, all methods 
are impacted, but methods based on rank correlations remain far superior.
All results and further details are available in the supplementary materials.

%% file: Sections/Application1.tex
\subsection{Application 1: Isoprenoid Synthesis in \emph{Arabidopsis thaliana}}
\label{application1}
\begin{figure}
    \centering
    \begin{center}
     \begin{subfigure}[t]{0.03\textwidth}
    \textbf{a)}
    \end{subfigure}
    \begin{subfigure}[t]{0.45\textwidth}
        \centering
        \adjustbox{valign=t}{
        \resizebox{\textwidth}{!}{
            \input{Figures/Application1/original_amat.tex}}
        }
        \caption{}
    \end{subfigure}\hfill
         \begin{subfigure}[t]{0.03\textwidth}
    \textbf{b)}
    \end{subfigure}
    \begin{subfigure}[t]{0.45\textwidth}
        \centering
            \adjustbox{valign=t}{\includegraphics[width=\textwidth]{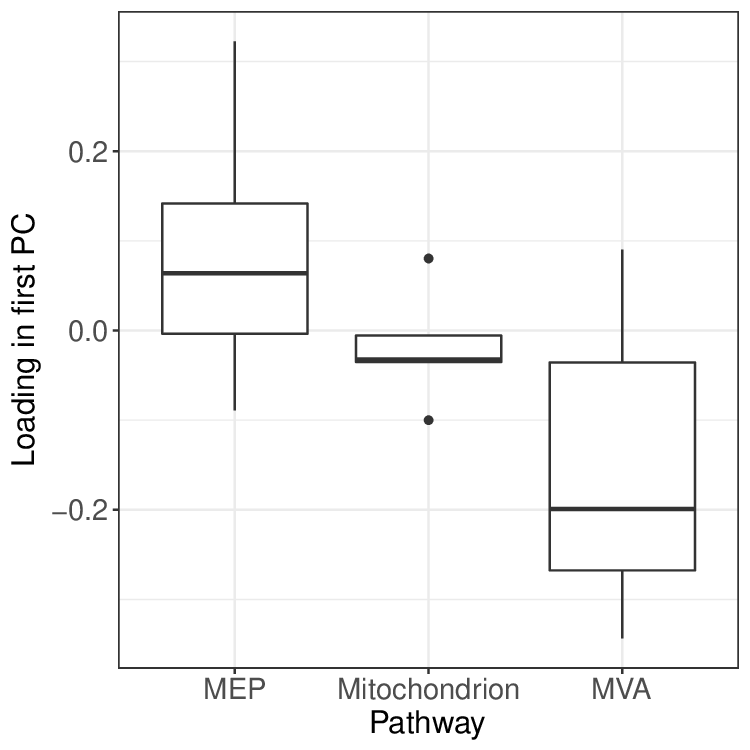}}
        \caption{}
    \end{subfigure}
    \caption{\textbf{a)} Directed graph induced by the MVA and MEP metabolic pathways, as shown in Figure 3 of \cite{Wille2004}. A non-zero entry (i,j) indicates a directed edge $i\to j$.
    \textbf{b)} Per pathway distribution of the entries of $\hat{L}$'s first eigenvector.
    }
    \label{App1Fig1}
    \end{center}
\end{figure}

We illustrate a few properties of our approach on a dataset containing
gene expression measurements taken in \emph{Arabidopsis thaliana} grown under $n=118$ different conditions (such as light/darkness, growth hormones, etc...) \citep{Wille2004}.
\cite{Wille2004} gave particular attention to the genes involved in isoprenoid synthesis.
In \emph{Arabidopsis thaliana}, two pathways, located in distinct organs, are responsible for isoprenoid synthesis: 
the mevalonate pathway (MVA) and the non-mevalonate pathway (MEP).
We downloaded the data made available in the supplementary materials
of \cite{Wille2004} and took the $p=33$ genes represented in Figure 3 of  \cite{Wille2004}.
They fall into three categories: genes that are part of the MVA pathway, 
genes that are in the MEP pathway and mitochondrial genes. For illustration, Figure \ref{App1Fig1} \textbf{a)}
shows the adjacency matrix of the metabolic pathways. 
This is a graph in which  nodes are genes and edges are chemical reactions between
gene products.
 This graph, while related
to the regulatory network we aim to estimate, is very different from it: in general, both the structure and the direction of the edges differ. 
However, it gives information about which genes are in which pathways.

We fitted LRpS and Kendall-LRpS to our data and selected the tuning parameters $\eta_n$ and $\gamma$ with five-fold cross-validation.
The low-rank matrix $\hat{L}$ estimated by LRpS had two non-zero eigenvalues, with ratio $\frac{\hat{\sigma}_1}{\hat{\sigma}_2} = 347$.
Hence, only the first eigenvector was retained.
In order to see whether we could interpret the hidden variables estimated by LRpS, we looked 
at the loadings of the genes in the first eigenvector of $\hat{L}$. Figure \ref{App1Fig1} \textbf{b)} shows the 
distribution of the loadings per pathway and suggests that the main source of variation
in the data is given by these pathways, which are sometimes unknown in less studied organisms.
By applying GES to ${\hat{K}_O}^{-1}$ -- the inverse of the sparse output of LRpS -- we are therefore modelling a regulatory network conditionally on those pathways, without 
having to provide further information.
Similar results were obtained with Kendal-LRpS+GES and are plotted 
in the appendix.

\begin{figure}
    \centering
    \begin{center}
     \begin{subfigure}[t]{0.01\textwidth}
    \textbf{a)}
    \end{subfigure}
    \begin{subfigure}[t]{0.28\textwidth}
        \centering
        \adjustbox{valign=t}{
        \resizebox{\textwidth}{!}{
            \input{Figures/Application1/GES_amat.tex}}
        }
        \caption{}
    \end{subfigure}\hfill
    \begin{subfigure}[t]{0.01\textwidth}
    \textbf{b)}
    \end{subfigure}
    \begin{subfigure}[t]{0.28\textwidth}
        \centering
        \adjustbox{valign=t}{
        \resizebox{\textwidth}{!}{
            \input{Figures/Application1/LRPS_GES_amat.tex}}
        }
        \caption{}
    \end{subfigure}
    \begin{subfigure}[t]{0.08\textwidth}
        \centering
        \adjustbox{valign=t}{
        \resizebox{\textwidth}{!}{
        \includegraphics{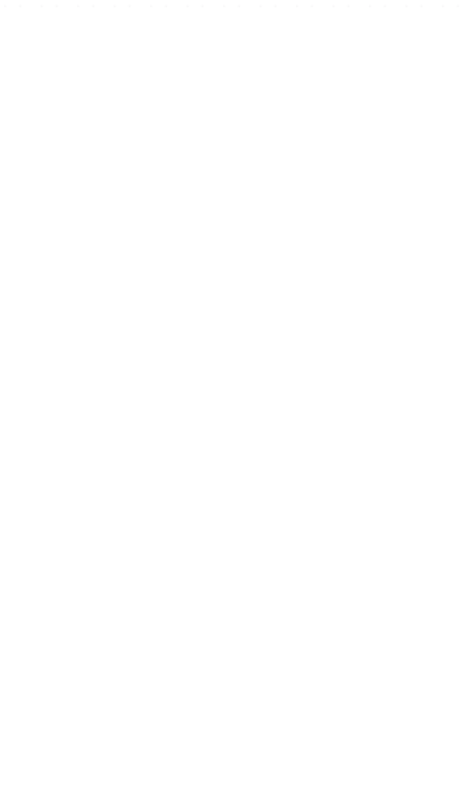}
        }}
        \caption{}
    \end{subfigure}\hfill
         \begin{subfigure}[t]{0.01\textwidth}
    \textbf{c)}
    \end{subfigure}
    \begin{subfigure}[t]{0.28\textwidth}
        \centering
        \adjustbox{valign=t}{
        \resizebox{\textwidth}{!}{
            \input{Figures/Application1/RFCI_amat.tex}}
        }
        \caption{}
    \end{subfigure}\\
    \begin{subfigure}[t]{0.01\textwidth}
    \textbf{d)}
    \end{subfigure}
    \begin{subfigure}[t]{0.28\textwidth}
        \centering
        \adjustbox{valign=t}{
        \resizebox{\textwidth}{!}{
            \input{Figures/Application1/GES_CE.tex}}
        }
        \caption{}
    \end{subfigure}\hfill
    \begin{subfigure}[t]{0.01\textwidth}
    \textbf{e)}
    \end{subfigure}
    \begin{subfigure}[t]{0.28\textwidth}
        \centering
        \adjustbox{valign=t}{
        \resizebox{\textwidth}{!}{
            \input{Figures/Application1/LRPS_GES_CE.tex}}
        }
        \caption{}
    \end{subfigure}
    \begin{subfigure}[t]{0.08\textwidth}
        \centering
        \adjustbox{valign=t}{
        \resizebox{\textwidth}{!}{
        \includegraphics{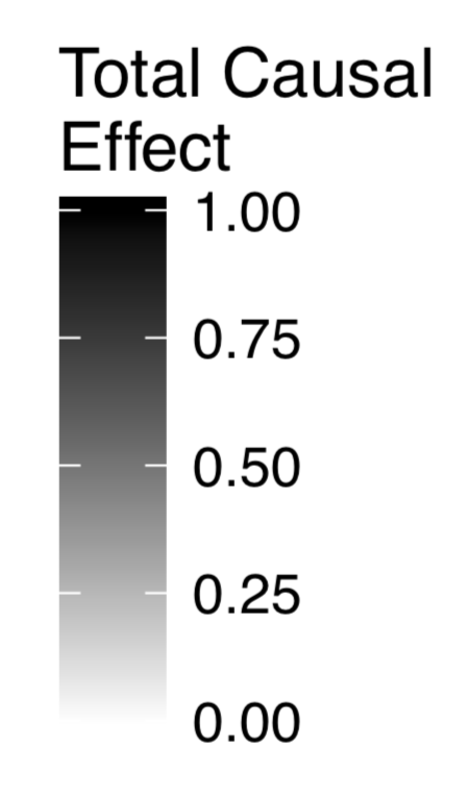}
        }}
        \caption{}
    \end{subfigure}\hfill
    \begin{subfigure}[t]{0.01\textwidth}
        \textbf{f)}
    \end{subfigure}
    \begin{subfigure}[t]{0.28\textwidth}
        \centering
        \adjustbox{valign=t}{
        \resizebox{\textwidth}{!}{
            \input{Figures/Application1/RFCI_CE.tex}}
        }
        \caption{}
    \end{subfigure}
    \caption{Estimates obtained by applying GES, LRpS+GES and RFCI to the data of \cite{Wille2004}.
    In the top row, an entry in the $i$th row and $j$th column indicates an arrow, tail or circle mark from the gene labelled by the $i$th row to 
    the gene labelled by the $j$th column.
    Edgemarks are as follows: circles are red, tails are blue, arrowheads are black.
    In bottom row, an entry in the $i$th row and $j$th column indicates a non-zero total causal effect from the gene labelled by the $i$th row to the gene labelled by the $j$th column.
    \textbf{a)} Adjacency matrix of the CPDAG estimated by GES. \textbf{b)} As in a), but with LRpS+GES. \textbf{c)} 
    Adjacency matrix of the PAG
    estimated by RFCI with $\alpha = \frac{0.5}{\sqrt{n}}$.  \textbf{d)} Matrix of total causal effects for GES,IDA. \textbf{e)} As in d), but with LRpS+GES.
    \textbf{f)} Matrix of total causal effects for RFCI,LV-IDA.
    }
    \label{App1Fig2}
    \end{center}
\end{figure}

In Figures \ref{App1Fig2} \textbf{a)} and \textbf{b)}, we show the adjacency matrices of the CPDAGs obtained by running GES and LRpS+GES
using the BIC score for GES.
Figure \ref{App1Fig2} \textbf{c)} shows the adjacency matrix of the PAG obtained by RFCI, with $\alpha = \frac{0.5}{\sqrt{n}}$ as before.
In Figures \ref{App1Fig2} \textbf{d)}, \textbf{e)}, the matrices of total causal effects computed from GES,IDA and (LRPS+GES),IDA are plotted, where IDA is used as in our simulations. 
In Figure \ref{App1Fig2} \textbf{f)}, the output of LV-IDA is plotted, with NAs marked in red.
The graphs and total causal effects obtained from other methods (NSDIST,IDA, etc\ldots) are plotted in the appendix. 
The output of Kendall-LRpS+GES is also shown in the appendix and 
differs from LRpS+GES in that there are more circle marks and slightly 
fewer edges. Qualitatively, it yields results that are similar to 
LRpS+GES.
Figure \ref{App1Fig2} illustrates the tendency of LV-IDA to produce very conservative estimates of the causal effects, with many pairs being either zero or NA.
On the other extreme, the causal effects of GES are stronger than those of LRpS+GES.
In particular, LRpS+GES does not find any strong causal relationship between mitochondrial genes and any other genes, as indicated by the ``white cross'' in the middle
of the matrix plotted in Figure \ref{App1Fig2} \textbf{e)}.
Both GES and LRpS+GES support the hypothesis of cross-talk from the MEP to the MVA pathway.

The metabolic pathways of \emph{Arabidopsis thaliana} have been studied in detail but, to the best of our knowledge, no reliable ground-truth is available for its directed regulatory network. 
For that reason, it is difficult to assess the quality of the estimated CPDAGs or matrices of total causal effects.
Nonetheless, we were able to show that the various methods
can yield very different results and to qualitatively assess them.
This application also gave us the opportunity to compare LV-IDA to other IDA-based methods on a real dataset. 

%% file: Figures/Application1/original_amat.tex
\begin{tikzpicture}[scale=1]
\node at (0,0) {\includegraphics[scale=1]{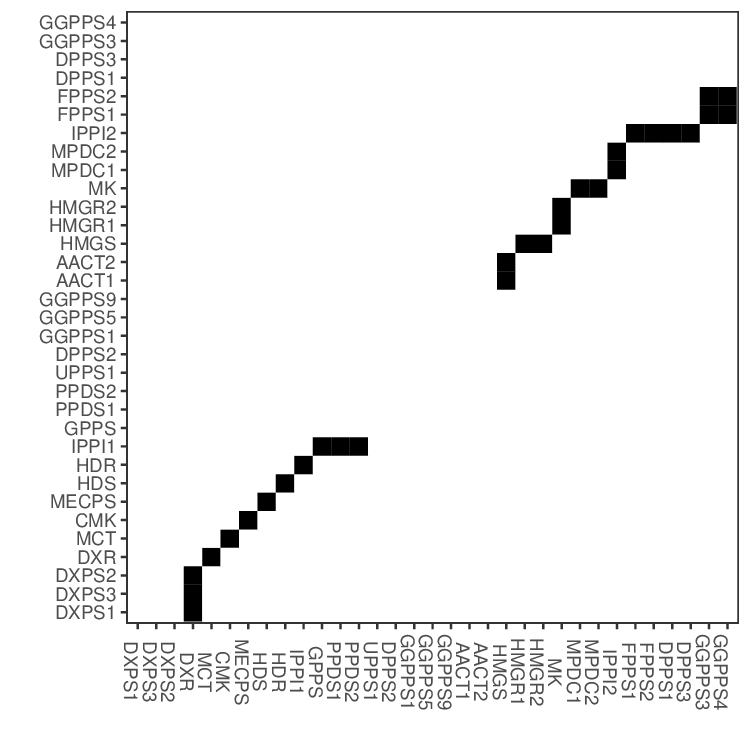}};
\draw [decorate,decoration={brace,amplitude=10pt,raise=4pt},thick]
(-0.2,-5.6) -- (-4.3,-5.6) node [black,midway,yshift=-1cm] {\textbf{MEP}};
\draw [decorate,decoration={brace,amplitude=10pt,raise=4pt},thick]
(1.3,-5.6) -- (-0.1,-5.6) node [black,midway,yshift=-1cm] {\textbf{Mit.}};
\draw [decorate,decoration={brace,amplitude=10pt,raise=4pt},thick]
(6,-5.6) -- (1.4,-5.6) node [black,midway,yshift=-1cm] {\textbf{MVA}};

\end{tikzpicture}

%% file: Figures/Application1/GES_amat.tex
\begin{tikzpicture}[scale=1]
\node at (0,0) {\includegraphics[scale=1]{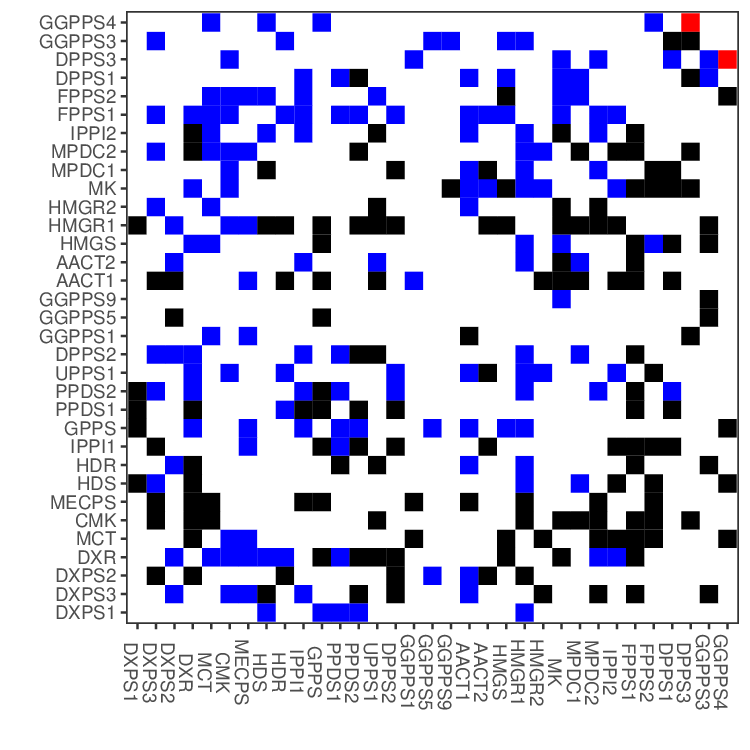}};
\draw [decorate,decoration={brace,amplitude=10pt,raise=4pt},thick]
(-0.2,-5.6) -- (-4.3,-5.6) node [black,midway,yshift=-1cm] {\textbf{MEP}};
\draw [decorate,decoration={brace,amplitude=10pt,raise=4pt},thick]
(1.3,-5.6) -- (-0.1,-5.6) node [black,midway,yshift=-1cm] {\textbf{Mit.}};
\draw [decorate,decoration={brace,amplitude=10pt,raise=4pt},thick]
(6,-5.6) -- (1.4,-5.6) node [black,midway,yshift=-1cm] {\textbf{MVA}};

\end{tikzpicture}

%% file: Figures/Application1/LRPS_GES_amat.tex
\begin{tikzpicture}[scale=1]
\node at (0,0) {\includegraphics[scale=1]{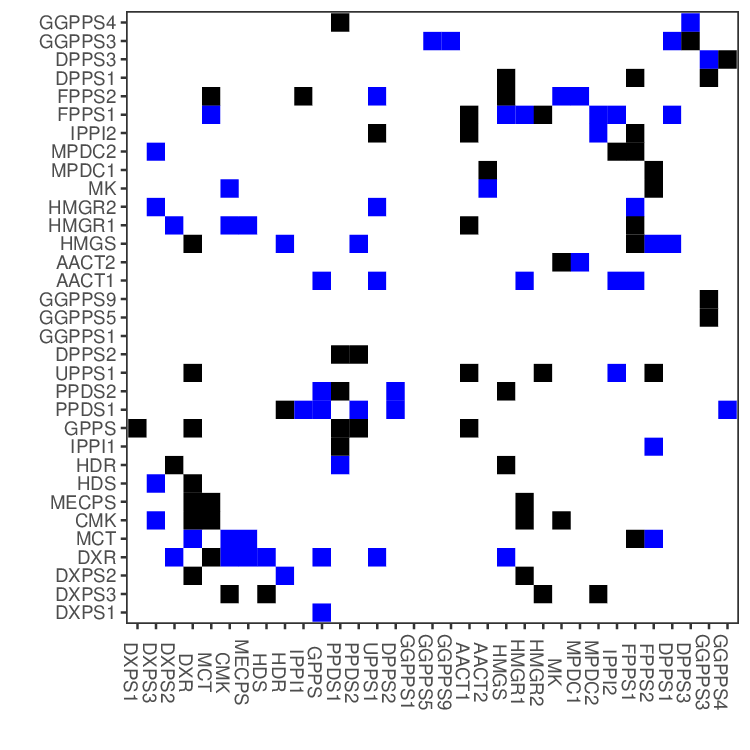}};
\draw [decorate,decoration={brace,amplitude=10pt,raise=4pt},thick]
(-0.2,-5.6) -- (-4.3,-5.6) node [black,midway,yshift=-1cm] {\textbf{MEP}};
\draw [decorate,decoration={brace,amplitude=10pt,raise=4pt},thick]
(1.3,-5.6) -- (-0.1,-5.6) node [black,midway,yshift=-1cm] {\textbf{Mit.}};
\draw [decorate,decoration={brace,amplitude=10pt,raise=4pt},thick]
(6,-5.6) -- (1.4,-5.6) node [black,midway,yshift=-1cm] {\textbf{MVA}};

\end{tikzpicture}

%% file: Figures/Application1/RFCI_amat.tex
\begin{tikzpicture}[scale=1]
\node at (0,0) {\includegraphics[scale=1]{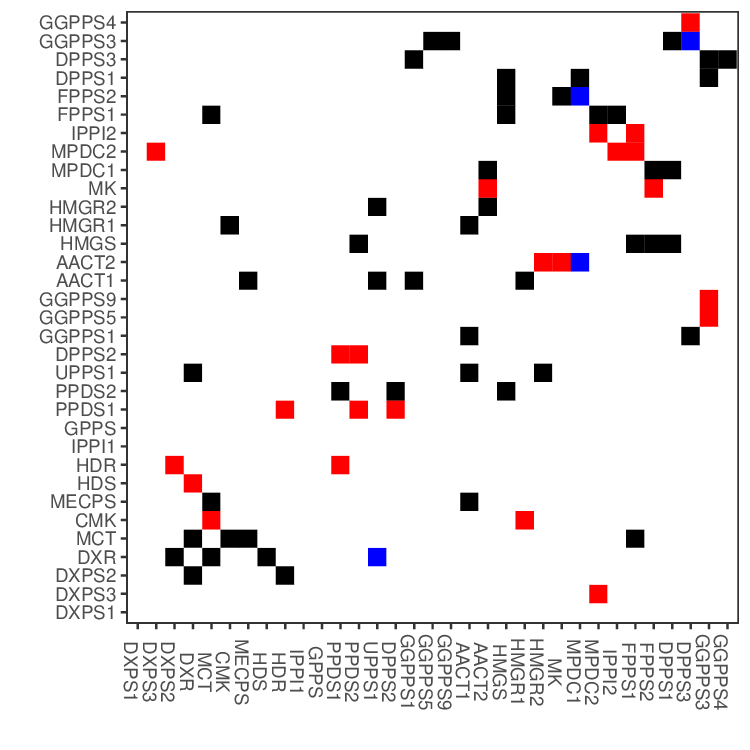}};
\draw [decorate,decoration={brace,amplitude=10pt,raise=4pt},thick]
(-0.2,-5.6) -- (-4.3,-5.6) node [black,midway,yshift=-1cm] {\textbf{MEP}};
\draw [decorate,decoration={brace,amplitude=10pt,raise=4pt},thick]
(1.3,-5.6) -- (-0.1,-5.6) node [black,midway,yshift=-1cm] {\textbf{Mit.}};
\draw [decorate,decoration={brace,amplitude=10pt,raise=4pt},thick]
(6,-5.6) -- (1.4,-5.6) node [black,midway,yshift=-1cm] {\textbf{MVA}};

\end{tikzpicture}

%% file: Figures/Application1/GES_CE.tex
\begin{tikzpicture}[scale=1]
\node at (0,0) {\includegraphics[scale=1]{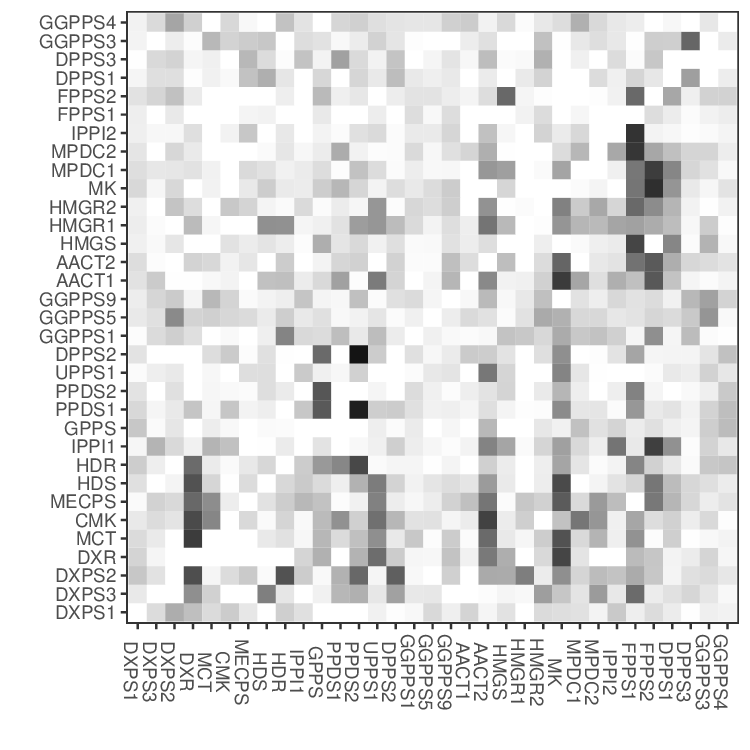}};
\draw [decorate,decoration={brace,amplitude=10pt,raise=4pt},thick]
(-0.2,-5.6) -- (-4.3,-5.6) node [black,midway,yshift=-1cm] {\textbf{MEP}};
\draw [decorate,decoration={brace,amplitude=10pt,raise=4pt},thick]
(1.3,-5.6) -- (-0.1,-5.6) node [black,midway,yshift=-1cm] {\textbf{Mit.}};
\draw [decorate,decoration={brace,amplitude=10pt,raise=4pt},thick]
(6,-5.6) -- (1.4,-5.6) node [black,midway,yshift=-1cm] {\textbf{MVA}};

\end{tikzpicture}

%% file: Figures/Application1/LRPS_GES_CE.tex
\begin{tikzpicture}[scale=1]
\node at (0,0) {\includegraphics[scale=1]{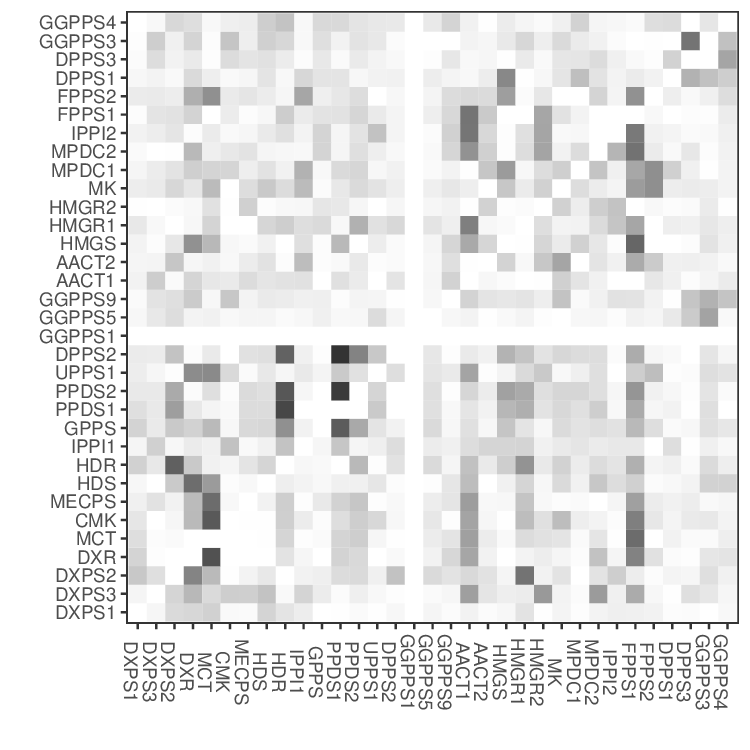}};
\draw [decorate,decoration={brace,amplitude=10pt,raise=4pt},thick]
(-0.2,-5.6) -- (-4.3,-5.6) node [black,midway,yshift=-1cm] {\textbf{MEP}};
\draw [decorate,decoration={brace,amplitude=10pt,raise=4pt},thick]
(1.3,-5.6) -- (-0.1,-5.6) node [black,midway,yshift=-1cm] {\textbf{Mit.}};
\draw [decorate,decoration={brace,amplitude=10pt,raise=4pt},thick]
(6,-5.6) -- (1.4,-5.6) node [black,midway,yshift=-1cm] {\textbf{MVA}};

\end{tikzpicture}

%% file: Figures/Application1/RFCI_CE.tex
\begin{tikzpicture}[scale=1]
\node at (0,0) {\includegraphics[scale=1]{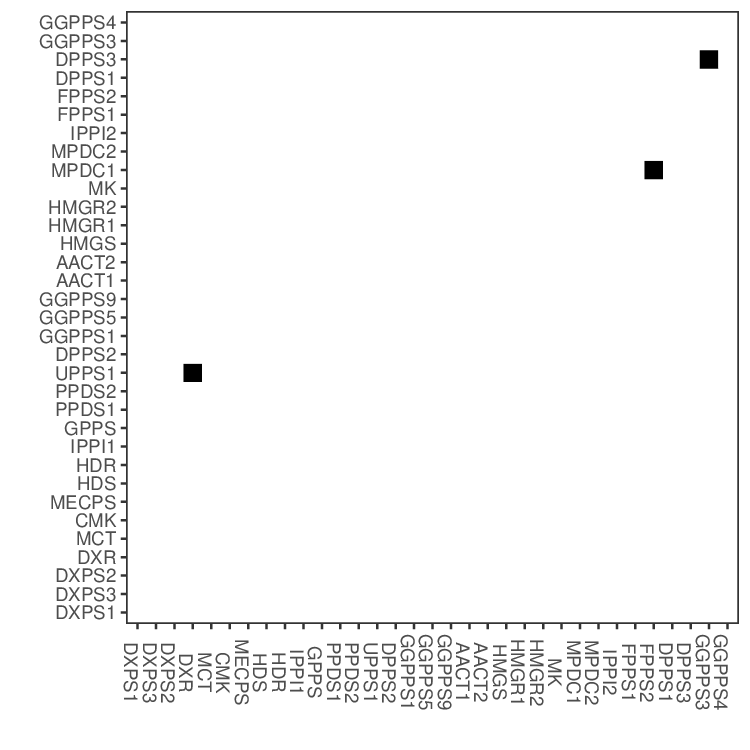}};
\draw [decorate,decoration={brace,amplitude=10pt,raise=4pt},thick]
(-0.2,-5.6) -- (-4.3,-5.6) node [black,midway,yshift=-1cm] {\textbf{MEP}};
\draw [decorate,decoration={brace,amplitude=10pt,raise=4pt},thick]
(1.3,-5.6) -- (-0.1,-5.6) node [black,midway,yshift=-1cm] {\textbf{Mit.}};
\draw [decorate,decoration={brace,amplitude=10pt,raise=4pt},thick]
(6,-5.6) -- (1.4,-5.6) node [black,midway,yshift=-1cm] {\textbf{MVA}};

\end{tikzpicture}

%% file: Sections/Application2.tex
\subsection{Application 2: Regulatory Network in Ovarian Cancer}

We now consider the  problem of identifying the targets regulated by a given set of transcription factors in a human gene expression
dataset. 
This problem is often considered in the literature because it constitutes
an example of a real-life dataset for which the existence and direction of \emph{some} edges is known, 
thus making it possible to compare estimated graphs to a ``partial ground-truth''  \citep{TsamardinosEtAl06, Han2016}. 
Briefly, a \emph{transcription factor} is a protein which regulates the mRNA expression of a gene by binding to a specific DNA sequence near
its promoting region.
Some families of  transcription factors have been studied in detail, and publicly available databases such as TRRUST provide
lists of transcription factors along with the genes -- called \emph{targets} -- they regulate \citep{Han2015}.
transcription factors play a crucial in role in cancer development, which is why it is believed that intervening on the expression of such genes
could alter the course of some cancers \citep{Darnell2002}.

In this application, we follow closely the steps described in Section 5.1 of \cite{Han2016} where
ovarian adenocarcinomas are studied. 
We used the RNA-Seq data available from the National Cancer Institute (\url{portal.gdc.cancer.gov/})
and log-transformed the gene expression levels. 
There is a consensus about how important some transcription factor families are for cancer development \citep{Darnell2002, Redell2005}. 
We therefore selected the transcription factors belonging to those families\footnote{Namely: FOS, FOSB, JUN, JUNB, JUND, ESR1, ESR2, AR, NFKB1, 
NFKB2, RELA, RELB, REL, STAT1, STAT2, STAT3, STAT4, STAT5, STAT6}.
Following \cite{Han2016}, we also extracted the genes that are known to have direct interactions with these transcription factors according to NetBox\footnote{\url{http://sanderlab.org/tools/netbox.html}}, ``a software tool for performing network analysis on human interaction networks which is pre-loaded with networks derived from four curated data sources, including the Human Protein Reference Database (HPRD), Reactome, NCI-Nature Pathway Interaction (PID) Database, and the MSKCC Cancer Cell Map''.
The resulting dataset contained $p = 501$ genes and $n = 247$ samples.

To construct a reference network to which we can compare our estimates, we used the output of NetBox.
NetBox outputs a list of known (unoriented) interactions between some of the 501 selected genes.
Unfortunately, nothing indicates
whether those interactions are causal; in general it is not because two genes interact in NetBox that intervening on the 
expression levels of one of the genes will induce a change in the expression level of the other.
However, thanks to our knowledge of transcription factors, 
we do know that whenever there is an interaction between a transcription factor and a non-transcription factor, then it is
likely to be causal and directed from the transcription factor to its target. 
Moreover, transcription factors are tissue specific, meaning that we can only expect a subset of the interactions to be active in
any given cell-type \citep{Eeckhoute09}.
These observations allow us to build three reference networks: 
\emph{a)} an undirected graph in which there is an
edge between A and B whenever they are said to interact according to NetBox (this is Network A);
\emph{b)} a ``causal'' undirected graph in which only edges between transcription factors and their targets have been retained (Network B);
\emph{c)} a causal directed graph in which the edges of Network B have been ordered from transcription factors to their targets (Network C).

In this application, the number of variables ($p=501$) is rather large compared to the sample size ($n=247$).
 We therefore selected the tuning parameters of LRpS ($\eta_n, \gamma$) using the Extended BIC 
instead of cross-validation \citep{FoygelDrton10}.

\begin{figure}
    \centering
    \begin{center}
     \begin{subfigure}[t]{0.03\textwidth}
    \textbf{a)}
    \end{subfigure}
    \begin{subfigure}[t]{0.42\textwidth}
        \centering
            \adjustbox{valign=t}{\includegraphics[width=\textwidth]{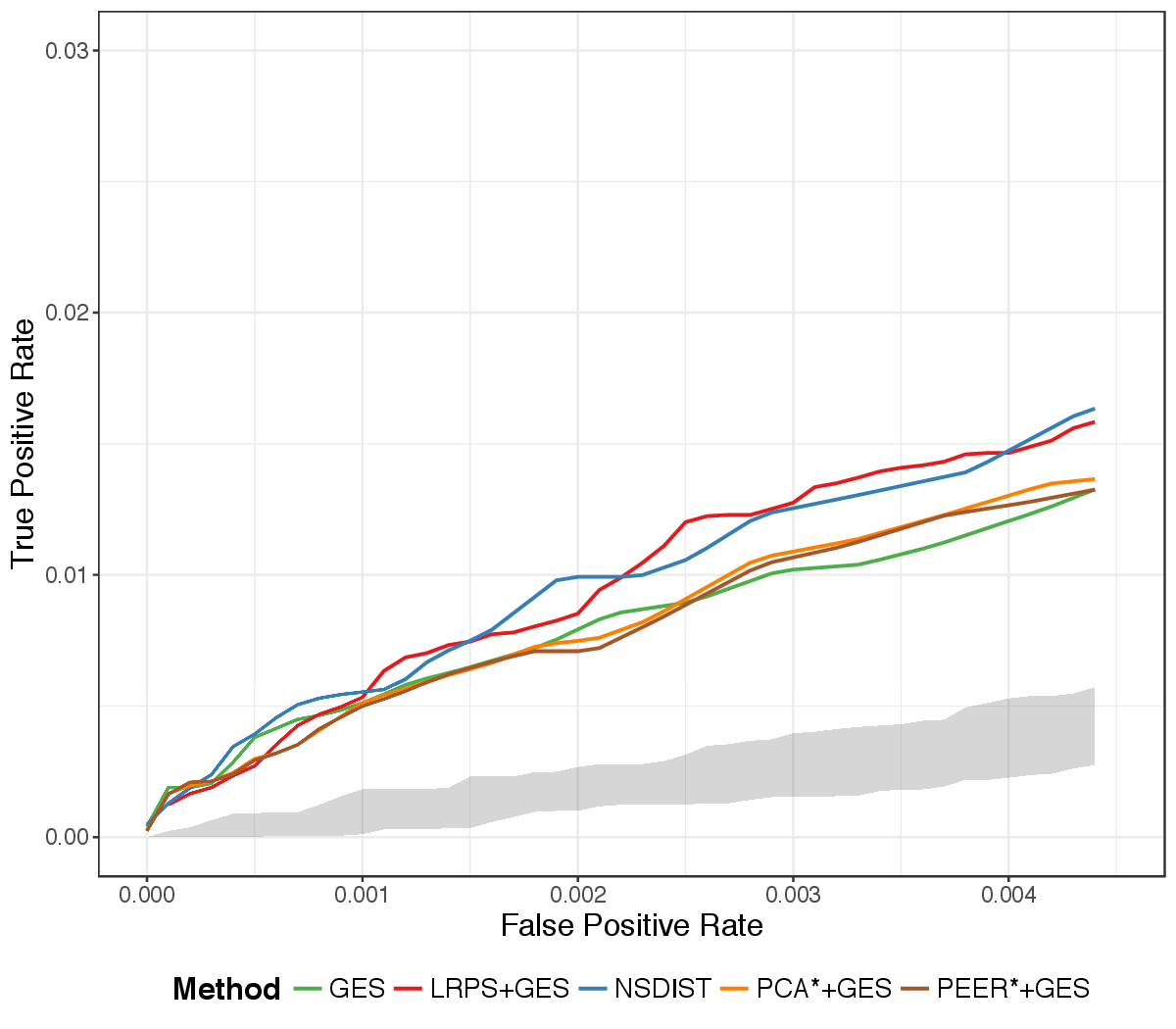}}
        \caption{}
    \end{subfigure}\hfill
         \begin{subfigure}[t]{0.03\textwidth}
    \textbf{b)}
    \end{subfigure}
    \begin{subfigure}[t]{0.42\textwidth}
        \centering
            \adjustbox{valign=t}{\includegraphics[width=\textwidth]{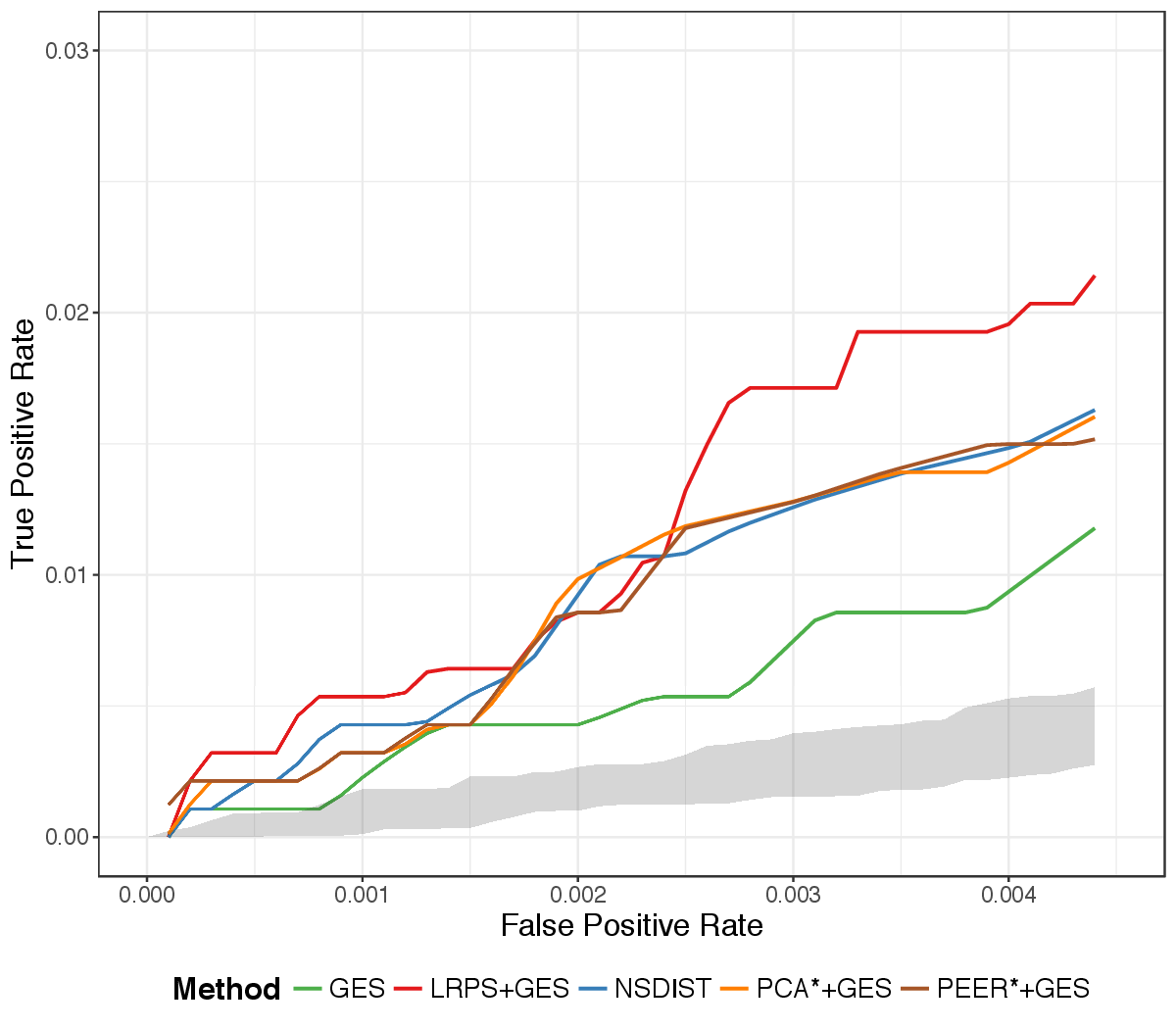}}
        \caption{}
    \end{subfigure}\\
    \begin{subfigure}[t]{0.03\textwidth}
    \textbf{c)}
    \end{subfigure}
    \begin{subfigure}[t]{0.42\textwidth}
        \centering
            \adjustbox{valign=t}{\includegraphics[width=\textwidth]{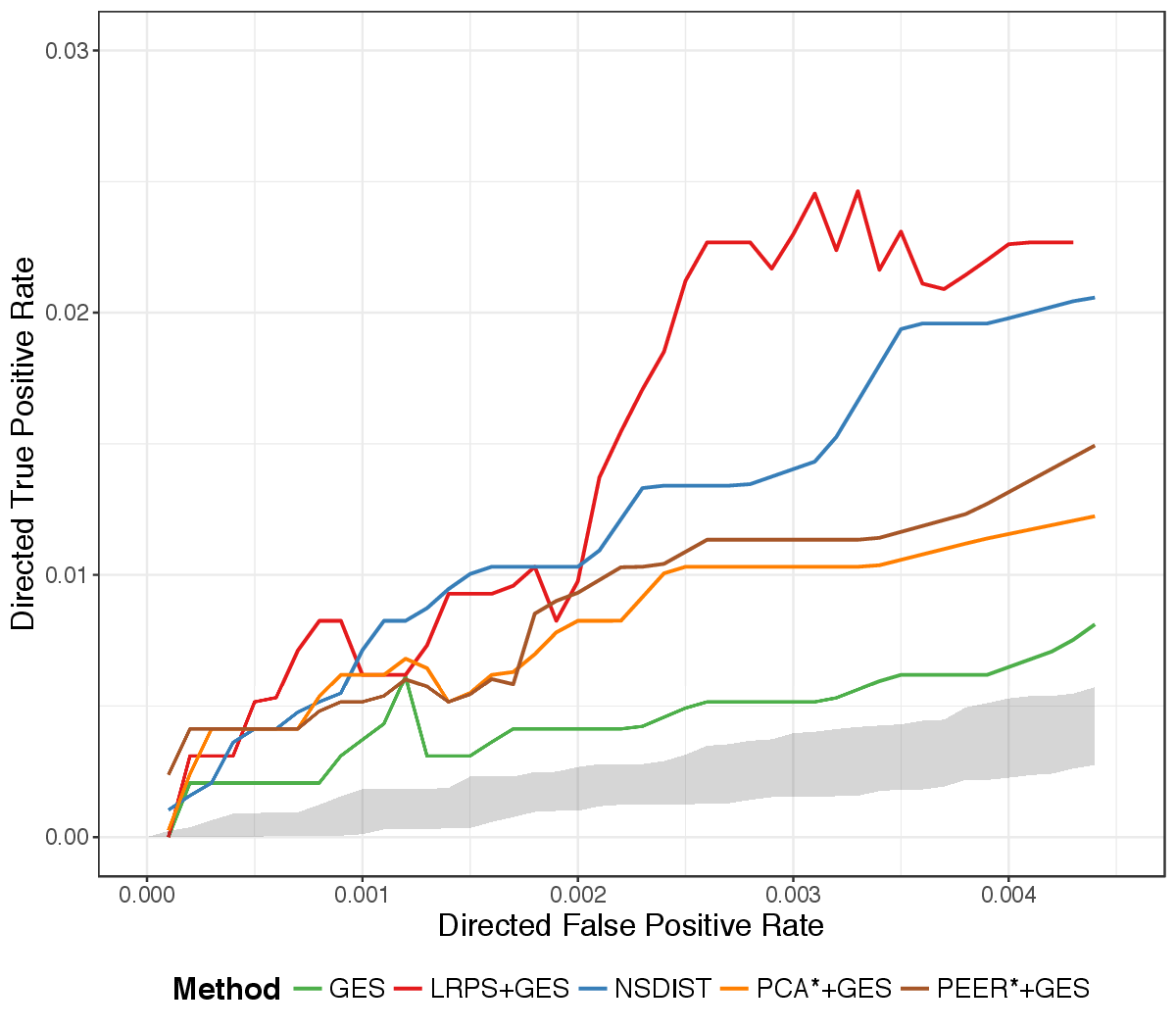}}
        \caption{}
    \end{subfigure}\hfill
         \begin{subfigure}[t]{0.03\textwidth}
    \textbf{d)}
    \end{subfigure}
    \begin{subfigure}[t]{0.42\textwidth}
        \centering
            \adjustbox{valign=t}{\includegraphics[width=\textwidth]{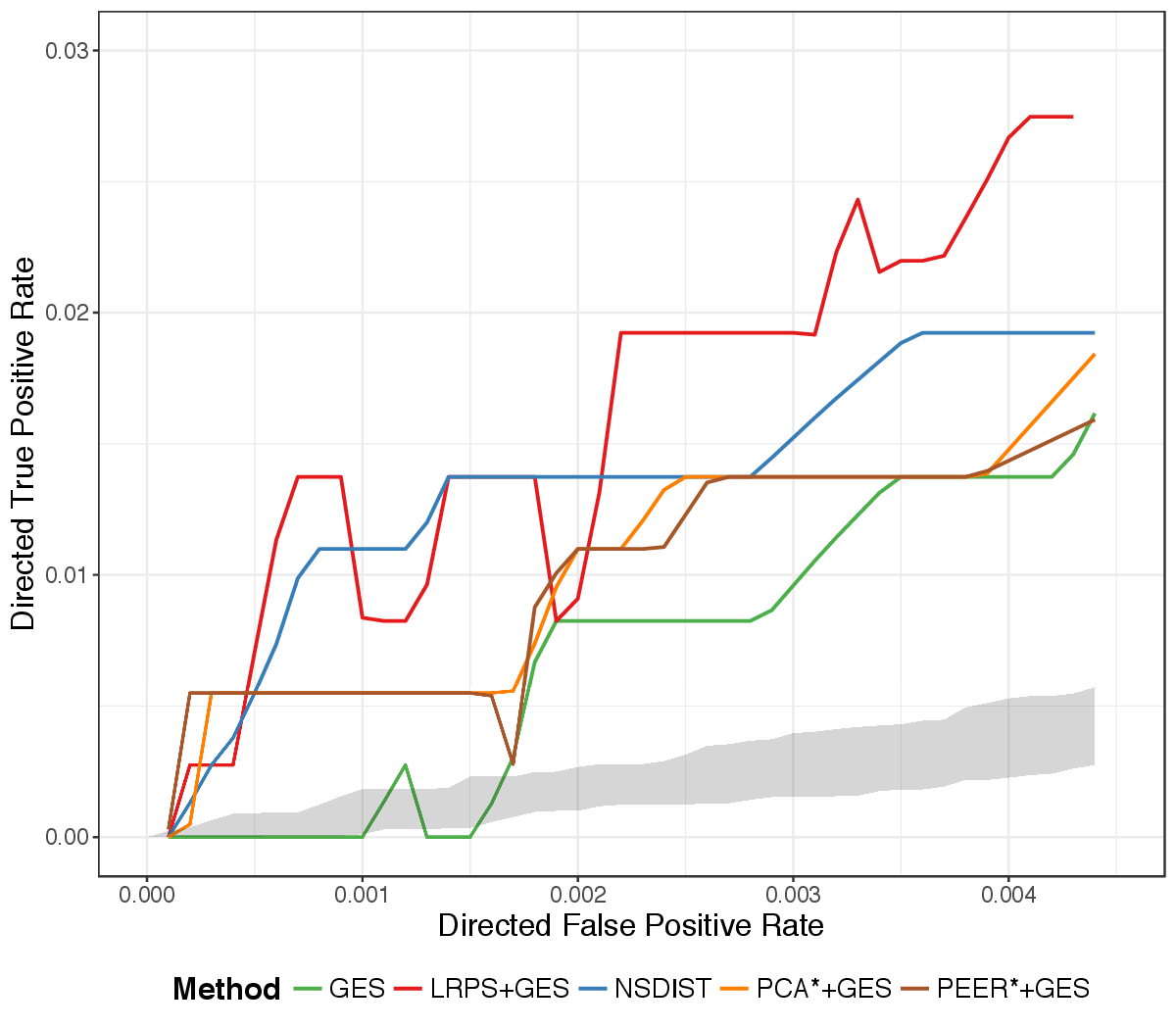}}
        \caption{}
    \end{subfigure}

    \caption{Comparison of the estimates of various methods against reference networks A, B, C and D. 
        \textbf{a)} ROC curves comparing the skeleton of the estimates to the undirected NetBox network (Network A).
        \textbf{b)} Same as a), but for the causal undirected network (Network B).
        \textbf{c)} ROC curves comparing the estimated CPDAG (or DAG in the case of NSDIST) to the directed causal network (Network C).
        \textbf{d)} Same as c), but with the directed network induced by TRRUST (Network D).
    }
    \label{App2Fig1}
    \end{center}
\end{figure}

 In Figure \ref{App2Fig1} we compare the output of various methods (GES, LRpS+GES, NSDIST, PCA*+GES, PEER*+GES) to reference networks
A, B and C in terms of True and False Positive Rates (TPR, FPR). 
For Network C, we follow again \cite{Han2016}: an undirected edge in a CPDAG is counted as half a true positive and half a false negative.
In grey, we plot the range spanned by the 2.5 and 97.5 percentiles of our null distribution.
It was computed by first picking 100 random ordering of the variables and, starting from a complete DAG, removing random edges one after the other until there are no edges left.
After each removal, we computed the performance metrics of the DAG with respect to all reference networks, thus 
generating 100 random regularisation paths for each of the plots.

Figure \ref{App2Fig1} \textbf{a)} plots the Receiving Operator Curve (ROC) for Network A. All methods display comparable performances, although 
NSDIST and LRpS+GES appear slightly above GES.
In Figure \ref{App2Fig1} \textbf{b)}, we restrict ourselves to Network B, so that only transcription factor-target edges are counted.
LRpS+GES is clearly above NSDIST, PCA*+GES and PEER*+GES which are themselves outperforming GES.
When the direction of the edges is also taken into account (Figure \ref{App2Fig1} \textbf{c)}), LRpS+GES remains ahead of the other
methods, and NSDIST beats PCA and PEER.
The difference in performance between NSDIST and GES does not come as a surprise since Figure \ref{App2Fig1} \textbf{a)} and \textbf{b)}
reproduce the findings of Figures 4 \textbf{a)} and \textbf{b)} of  \cite{Han2016}.

Since NetBox is not restricted to transcription factor-target interactions, we sought to confirm the results of Figure \ref{App2Fig1} \textbf{c)} by 
using an independent source of validation specialised in transcriptional regulatory relationships. 
We used TRRUST \citep{Han2015} and constructed Network D by adding directed edges between transcription factors and targets according
to TRRUST.
In Figure \ref{App2Fig1} \textbf{d)} we plot the resulting ROC cuve, which reproduces the results obtained 
using NetBox transcription factor-target interactions as ground truth.

Making definitive statements about the nature of the hidden confounders 
in this dataset is difficult. 
We can hypothesise that it is prone to the type of confounding 
typically seen in gene expression data where intersample 
heterogeneities (\emph{e.g.} relatedness, batch effects) are often responsible for unwanted variations. 
Gaining access to the patients' DNA would make it possible to test whether relatedness between samples is indeed a cause of confounding in our dataset. 
Batch effects can also be accounted for to some extent, but there will always remain confounders that cannot be ruled out. For example, it has been observed that factors as varied as the time postportem a sample is collected, or the ozone levels in the laboratory introduce spurious correlations \citep{Kang1909}. 

It is also possible that unobserved transcriptions factors, or transcription factors that are not in our database, are responsible 
for these gene-gene interactions. This highlights one of the limitations
of our method.

%% file: Sections/Discussion.tex
We discussed the problem of estimating the Markov equivalence class of a DAG in the presence of hidden variables. 
Building on previous work by \cite{ChandrasekaranParriloWillsky12} and \cite{Chickering02}, we suggested a two-stage approach -- termed LRpS+GES -- which first
removes unwanted variation using latent Gaussian graphical model selection, and then estimates a CPDAG by applying GES.
We chose GES for its good empirical performance and theoretical guarantees \citep{NandyHauserMaathuis16}, but we note that the second step can be replaced by any structure learning algorithm
for DAGs that assumes causal sufficiency -- although another choice
might not offer the same theoretical guarantees.
Our main theoretical result states that LRpS+GES is consistent for CPDAG recovery in some sparse high-dimensional regimes. 
Through simulations and two applications to gene expression datasets, we showed that our approach often outperforms 
the state of the art, both in terms of graphical structure recovery and total causal effect estimation.
Moreover, the results reported in our simulations can be achieved in practice since tuning parameter selection 
was performed using in-sample information only\footnote{The code for our simulations and applications is made available with this paper.}.
 
When it comes to removing unwanted variation from biological datasets, 
state-of-the-art approaches usually incorporate external information into the analysis by including additional covariates (\emph{e.g.} gender, genetic relatedness), 
thus also accounting for known confounders \citep{Stegle2012, Mostafavi2013}. 
Since these additional covariates are often discrete, modelling them with LRpS+GES would be a violation of our assumptions.
In such a setting, it is straightforward to replace LRpS by the LSCGGM estimator suggested in \cite{FrotEtAl18}. LSCGGM 
makes it possible to perform a low rank plus sparse decomposition, conditionally on a number of arbitrarily distributed random variables.
The LRpS+GES approach could therefore be replaced by the ``LSCGGM+GES'' estimator, which would come with similar theoretical 
guarantees.

The computational cost of LRpS+GES might also be a concern to the practitioner. 
In Algorithm \ref{alg1} we first estimate an inverse covariance matrix $\hat{K}_O$. 
To the best of our knowledge, the fastest algorithm for this LRpS step
uses the so-called alternative direction method  of multipliers, with a cost of $\mathcal{O}(p^3)$ per iteration \citep{Ma:2013}. 
Next, $\hat{K}_O$ must be inverted, at a cost of $\mathcal{O}(p^3)$, and then GES is run on $\hat{K}_O^{-1}$. 
For large problems, this last step can be replaced by the ARGES algorithm suggested in \citet{NandyHauserMaathuis16}. 

As detailed earlier, there exist other approaches which are capable of estimating DAG models and total causal effects 
in the presence of hidden variables, \emph{i.e.} FCI-type algorithms \citep{SpirtesMeekRichardson95, ColomboEtAl12,ClaassenMooijHeskes13} and LV-IDA \citep{Malinsky2017}.
In both our simulations and our first application, we found that  such approaches are very conservative under our assumptions. 
However, they do outperform LRpS+GES when hidden variables act on the observed ones in a sparse fashion.
As such, LRpS+GES is complementary to existing methods.

Finally, we note that the LRpS+GES estimator can be modified to tackle another widespread problem: selection bias.
Reusing the notations introduced in Section \ref{setup}, selection bias can be handled as follows. 
Let $X \in \mathbb{R}^{p+h}$ be a zero-mean random vector which follows a multivariate normal distribution with \emph{covariance} 
matrix 
$\begin{pmatrix} \Sigma_O^\ast & \Sigma_{OH}^\ast \\ {\Sigma_{OH}^\ast}^T& \Sigma_H^\ast \end{pmatrix}$. 
Let us further assume that there exists a DAG, $\mathcal{G}^\ast_O$ say, which is a perfect map of the distribution 
$\mathcal{N}(0, \Sigma_O^\ast)$.
Then, assuming that the variables in $X_H$ are selection variables, we only see observations from
\[
	X_O | X_H \sim \mathcal{N}\left(0, \Sigma_O^\ast - {\Sigma_{OH}^\ast} {\Sigma_H^\ast}^{-1} {\Sigma_{OH}^{\ast T}} \right).
\] 
By the Woodbury identity, this can be rewritten in terms of the precision matrix as
\[
	X_O | X_H \sim \mathcal{N}\left(0, \left({\Sigma_O^\ast}^{-1} - L^{\ast}\right)^{-1}\right),
\] 
where $L^\ast$ is a \emph{negative semi-definite} matrix defined as
\[ L^\ast := -{\Sigma_O^\ast}^{-1} {\Sigma_{OH}^\ast} \left(\Sigma_H^\ast -   {\Sigma_{OH}^{\ast T}} {\Sigma_O^\ast}^{-1} {\Sigma_{OH}^\ast} \right)^{-1}{\Sigma_{OH}^{\ast T}} {\Sigma_O^\ast}^{-1}.\]
Since Theorem 4.1 of \cite{ChandrasekaranParriloWillsky12} does not make any 
assumptions about the positive-definitiveness of $L^\ast$, the following estimator could replace LRpS in the first stage of LRpS+GES: 
\begin{equation}\label{lrps_estimator2}
    \argmin_{K_O - L \succ 0, L \preceq 0} -\ell(K_O - L; \Sigma^n_O) + \eta_n(\gamma ||K_O||_1 + ||L||_\ast),
\end{equation}
where $\ell(K;\Sigma^n_O) = -\trace(K \Sigma_O^n) + \log \det K$ and $\eta_n,\gamma > 0$. 
This modified approach is consistent in the presence of selection variables under the similar conditions as Theorem \ref{theo1}. 
The only difference is in the interpretation of the 
condition $\xi(T)\mu(\Omega) \leq \frac{1}{6} C^2$, which would require that ${\Sigma_O^\ast}^{-1}$ be sparse and that there be 
few selection variables that are directly regulated by many of the observed variables.